\documentclass[letterpaper, 11pt]{article}

\usepackage{mathtools}
\usepackage{booktabs}
\usepackage{array}
\usepackage{enumitem}
\usepackage{amsmath,amsfonts,amssymb,bm}
\usepackage{titlesec}
\usepackage{pgfgantt,rotating}
\usepackage{float}
\usepackage{graphicx}
\usepackage{sidecap}
\usepackage{wrapfig}
\usepackage{bold-extra}
\usepackage{authblk}
\usepackage{appendix}
\usepackage{makecell}
\usepackage{dirtytalk}
\usepackage{placeins}
\usepackage[export]{adjustbox}
\usepackage{cite}

\usepackage[margin=1in]{geometry}

\newcommand{\beq}{\begin{equation}}
\newcommand{\eeq}{\end{equation}}
\newcommand{\bgqar}{\begin{eqnarray}}
\newcommand{\enqar}{\end{eqnarray}}
\newcommand{\bgqarn}{\begin{eqnarray*}}
\newcommand{\enqarn}{\end{eqnarray*}}
\newcommand{\bgary}{\begin{array}}
\newcommand{\enary}{\end{array}}

\graphicspath{{Figures/}}

\title{Vision-Based Structural Damage Identification in Vibrating Beams via Dynamic Mode Decomposition}
\author{R K B M Rizmi}
\author{Shabbir Ahmed \footnote{Corresponding author.}}

\affil{\small Dynamical Systems and Signals Lab (DSSL)  \\ Department of Mechanical Engineering \\ South Dakota State University, Brookings, SD \\ Email: \{rkbm.rizmi,shabbir.ahmed\}@sdstate.edu }

\date{\today}

\begin{document}

\maketitle

%------------------------------------------------------------------------------------------------------------
% Abstract
%------------------------------------------------------------------------------------------------------------

\begin{abstract}

Structural damage detection using non-contact sensing remains a challenging problem in structural health monitoring. This study presents a data-driven framework based on Dynamic Mode Decomposition (DMD) for extracting structural dynamics directly from high-speed video recordings of vibrating structures. Within this approach, the underlying dynamics are approximated by a linear operator, whose spectral decomposition yields modal frequencies and corresponding spatial mode shapes, enabling a physically interpretable representation of the system response. The proposed methodology is evaluated through both numerical and experimental investigations. First, a cantilever beam model is simulated in ANSYS under healthy and damaged conditions. DMD is applied to partial observation data to reconstruct and predict the system response, while the extracted modal features are analyzed to characterize damage-induced variations. Second, high-speed video recordings of the beam are processed into spatiotemporal snapshot matrices, allowing DMD to recover full-field dynamic behavior without contact sensors. To enable quantitative assessment, a damage index is formulated based on DMD-derived modal features, capturing deviations in both frequency content and spatial characteristics. The results demonstrate consistent and distinguishable patterns between healthy and damaged states across both simulation and experiments, highlighting the capability of DMD as a robust and interpretable tool for non-contact damage detection using video data.
\vspace{0.3cm}


\end{abstract}

%------------------------------------------------------------------------------------------------------------
%                              Important conventions and symbols -- Acronyms
%------------------------------------------------------------------------------------------------------------

%------------------------------------------------------------------------------------------------------------

\newpage\pagebreak 

\tableofcontents 
\newpage\pagebreak
%------------------------------------------------------------------------------------------------------------
% Introduction
%------------------------------------------------------------------------------------------------------------
\section*{Nomenclature}

\begin{tabular}{ll}
$A$ & Cross-sectional area [m$^2$] \\
$c_d$ & Viscous damping coefficient per unit length [N\,s\,m$^{-2}$] \\
$C_k$ & Modal damping coefficient [kg\,s$^{-1}$] \\
$E$ & Young's modulus [Pa] \\
$I$ & Second moment of area [m$^4$] \\
$K_k$ & Modal stiffness [N\,m$^{-1}$] \\
$L$ & Beam length [m] \\
$M_k$ & Modal mass [kg] \\
$q_k(t)$ & Modal coordinate \\
$w(x,t)$ & Transverse displacement [m] \\
$\beta_k$ & Frequency parameter/eigenvalue [m$^{-1}$] \\
$\zeta_k$ & Modal damping ratio [-] \\
$\rho$ & Material density [kg\,m$^{-3}$] \\
$\phi_k(x)$ & Mode shape \\
$\omega_k$ & Natural frequency [rad\,s$^{-1}$] \\
$\omega_{dk}$ & Damped natural frequency [rad\,s$^{-1}$] \\
$k$ & Mode number \\
MS & Mode Shape \\
MSS & Mode Shape Slope \\
MSC & Mode Shape Curvature \\
MSCS & Mode Shape Curvature Square \\
\end{tabular}
\newpage\pagebreak
\section{Introduction} \label{sec:intro}
Structural health monitoring (SHM) is essential for ensuring the safety, reliability, and longevity of civil, mechanical, and aerospace systems \cite{Farrar-Worden07}. Among the various SHM strategies, vibration-based methods are widely used, as they characterize system behavior through modal properties such as natural frequencies, damping ratios, and mode shapes \cite{Kopsaftopoulos-Fassois13,Pandey1991,Wahab1999}. Structural damage often leads to stiffness degradation, which in turn alters these modal characteristics \cite{Pandey1991}. Consequently, changes in the dynamic response are commonly used as indicators of damage in vibration-based SHM \cite{Ahmed-Kopsaftopoulos19,ahmed-stationary}.

Experimental vibration characterization is traditionally performed using contact sensors such as accelerometers and strain gauges. While these sensors provide reliable measurements, they are inherently intrusive and offer limited spatial coverage, requiring dense sensor networks to capture distributed structural behavior. This limitation becomes particularly significant for large or complex structures, where sparse measurements may fail to capture localized damage. Additionally, sensor installation, wiring, and maintenance can increase system complexity and cost. Non-contact measurement techniques, such as laser vibrometry, address some of these challenges but often remain limited to point-wise measurements or require expensive instrumentation for full-field analysis \cite{Cheng2025_CVVibration}. In contrast, advances in video acquisition technology provide a promising alternative, enabling non-contact, full-field observation of structural vibrations in a cost-effective and scalable manner. However, extracting physically meaningful dynamic information from such high-dimensional visual data remains a key challenge.

Extracting modal information from video measurements necessitates methods capable of operating directly on high-dimensional spatiotemporal data. In this context, data-driven system identification provides a robust framework for modal analysis when explicit governing equations or accurate parametric models are unavailable. Among these, Dynamic Mode Decomposition (DMD) provides an operator-theoretic approach for approximating the Koopman operator, which governs the evolution of observables directly from sequential snapshots \cite{schmid2010dynamic}. By determining the best-fit linear operator that maps a data manifold to its subsequent state in time, DMD decomposes complex dynamics into physically meaningful modes \cite{brunton2022data}. These modes are associated with specific complex eigenvalues, which characterize the oscillatory frequencies and exponential growth or decay rates of the underlying dynamics. This facilitates a compact, dynamically consistent representation of high-dimensional systems. Originally developed for fluid dynamics, DMD’s efficacy in distilling coherent structural modes has led to its successful applications in structural mechanics and vibrations.

Prior works have focused on estimating the displacement field from image sequences. Saito and Kuno \cite{saito2020data} have applied DMD to experimental displacement field data of a cantilever beam obtained by video recordings, where they demonstrated that modal parameters such as natural frequencies and damping ratios can be extracted accurately. Garrido et al. \cite{garrido2023damage} have performed damage detection and localization of a cantilever beam from short burst video records by extracting the displacement field. Resende et al. \cite{resende2024damage} employed convolutional neural networks (CNNs) and displacement measurements for structural damage identification with data obtained from video recordings. However, relying on explicit displacement reconstruction necessitates intensive preprocessing and calibration, which often introduces artifacts and heightened measurement uncertainty. Furthermore, the decoupling of feature extraction from modal identification increases computational complexity and may obscure the underlying dynamics present in the raw pixel-intensity evolution.

An emerging paradigm in vision-based vibration characterization treats the video sequence itself as the fundamental system state, observing motion through an Eulerian framework where dynamics are captured via pixel-intensity fluctuations at fixed spatial coordinates. Colombo et al. \cite{colombo2025damage} proposed applying dynamic mode decomposition (DMD) directly to raw video frames. By treating individual pixel intensities as state measurements, this approach enables the direct computation of image dynamic modes, significantly reducing the preprocessing steps associated with displacement reconstruction.

However, this framework may introduce methodological complexities such as relying on optimized DMD (optDMD) rather than the standard approach and not extracting damage signatures directly from the raw modes. In this work, we propose to extract damage related features directly from the mode shapes and aim to formulate a novel damage index. Because structural damage manifests as perturbations in modal properties and mode shapes and these modes contain latent damage-sensitive signatures. Additionally, translating the visual variations of mode shapes into robust damage indicators across varying structural states or damage states necessitates the development of more interpretable and robust comparison metrics. Previously, mode shape curvature square was utilized for the detection and localization of damage in a beam structure from vibration response signal \cite{rucevskis2010damage, rucevskis2009multiple}. The method only required mode shape information and still provided damage information in a reliable way. In the context of ultrasonic guided wave-based damage detection and identification, Ahmed and Kopsaftopoulos have proposed a statistical damage diagnosis framework using a functional series time-dependent autoregressive (FS-TAR) model, where a probabilistic quantity termed as the "characteristic quantity" was utilized to detect and identify different damage levels \cite{ahmed2025functional}. In this work, we have formulated a novel characteristic quantity and damage index that may facilitate easier damage detection and identification using only the mode shape information. Additionally, we also utilized eigenvalue information extracted from the DMD model to formulate interpretable damage identification features and metrics. 

\section{Problem Statement}

The existing frameworks often rely on specialized variants like optimized DMD (optDMD) and multi-stage post-processing to isolate damage signatures. This presents an opportunity to develop more streamlined approaches that extract diagnostic features directly from the raw modal manifold. Since structural damage inherently manifests as perturbations in modal properties, the dynamic modes contain latent, damage-sensitive signatures that can be leveraged for health monitoring. Drawing inspiration from classical vibration-based methods, such as the use of mode shape curvature for damage localization \cite{rucevskis2010damage, rucevskis2009multiple}, and statistical diagnosis frameworks like the functional series time-dependent autoregressive (FS-TAR) models \cite{ahmed2025functional}, this work proposes a direct diagnostic methodology. We formulate a novel 'characteristic quantity' and damage index derived exclusively from DMD-extracted mode shapes and eigenvalues. This approach facilitates efficient damage detection and identification directly from the video data to interpretable damage metrics, bypassing the need for extensive heuristic pre-processing and post-processing.

\section{Theoretical Background}
\label{sec:method}
\subsection{Beam Theory}

The transverse vibration of slender beams is commonly described by the Euler-Bernoulli beam theory, which assumes small deflections, linear elasticity, and negligible shear deformation. For a uniform cantilever beam undergoing free vibration with distributed viscous damping, the governing equation can be written as

\begin{equation}
  \rho A\,\frac{\partial^2 w(x,t)}{\partial t^2}
  + c_d\,\frac{\partial w(x,t)}{\partial t}
  + EI\,\frac{\partial^4 w(x,t)}{\partial x^4}
  = 0,
  \label{eq:damped_pde}
\end{equation}

where $w(x,t)$ is the transverse displacement, $E$ is Young's modulus, $I$ is the second moment of area, $\rho$ is the density, $A$ is the cross-sectional area, and $c_d$ is a distributed viscous damping coefficient (force per unit length per unit velocity).

For a cantilever beam of length $L$, the boundary conditions are
\begin{align}
  w(0,t) &= 0, \quad \frac{\partial w}{\partial x}(0,t) = 0, \\
  \frac{\partial^2 w}{\partial x^2}(L,t) &= 0, \quad \frac{\partial^3 w}{\partial x^3}(L,t) = 0,
\end{align}
corresponding to a clamped-free configuration. The solution of Equation \ref{eq:damped_pde} can be expressed as a modal expansion as follows

\begin{equation}
  w(x,t) = \sum_{k=1}^{\infty} \phi_k(x)\,q_k(t),
  \label{eq:modal_expansion}
\end{equation}
where $\phi_k(x)$ are the eigenfunctions of the associated undamped spatial operator, and $q_k(t)$ are the modal coordinates. Substituting Eq.~\eqref{eq:modal_expansion} into the \emph{undamped} version of Eq.~\eqref{eq:damped_pde} and applying separation of variables yields the spatial eigenvalue problem

\begin{equation}
  EI\,\phi_k''''(x) = \rho A\,\omega_k^2\,\phi_k(x),
  \label{eq:eigen_problem}
\end{equation}

with the cantilever boundary conditions, where the wavenumber $\beta_k$ is defined through

\begin{equation}
  \beta_k^4 = \frac{\rho A\,\omega_k^2}{EI}.
  \label{eq:wavenumber}
\end{equation}
The corresponding eigenfunctions take the closed form

\begin{equation}
  \phi_k(x) = \cosh(\beta_k x) - \cos(\beta_k x)
  - \sigma_k\bigl(\sinh(\beta_k x) - \sin(\beta_k x)\bigr),
  \label{eq:eigenfunction}
\end{equation}

where

\begin{equation}
  \sigma_k = \frac{\cosh(\beta_k L)+\cos(\beta_k L)}{\sinh(\beta_k L)+\sin(\beta_k L)}.
\end{equation}

The wavenumbers $\beta_k$ are determined by the characteristic equation

\begin{equation}
  \cos(\beta_k L)\cosh(\beta_k L) + 1 = 0,
  \label{eq:characteristic_eq}
\end{equation}

and the corresponding natural frequencies are given by

\begin{equation}
  \omega_k = \beta_k^2 \sqrt{\frac{EI}{\rho A}}.
  \label{eq:natural_freq}
\end{equation}

For uniform $c_d$, the damping operator shares the eigenfunctions of the stiffness operator, ensuring the modal equations remain decoupled. Projecting Eq.~\eqref{eq:damped_pde} onto the mode shapes and exploiting this orthogonality yields a set of decoupled modal equations

\begin{equation}
  M_k\,\ddot{q}_k(t) + C_k\,\dot{q}_k(t) + K_k\,q_k(t) = 0,
  \label{eq:modal_mck}
\end{equation}

where

\begin{align}
  M_k &= \int_0^L \rho A\,\phi_k^2(x)\,dx, \\
  C_k &= \int_0^L c_d\,\phi_k^2(x)\,dx, \\
  K_k &= \int_0^L EI\,\left(\phi_k''(x)\right)^2 dx,
\end{align}

where the expression for $K_k$ follows from integration by parts of $\int_0^L \phi_k\, EI\,\phi_k''''\,dx$ and application of the free-end boundary conditions, which eliminate the boundary terms. Using the eigenvalue relation in Eq.~\eqref{eq:eigen_problem}, the modal stiffness can equivalently be written as

\begin{equation}
  K_k = M_k \omega_k^2.
\end{equation}

Normalizing Eq.~\eqref{eq:modal_mck} by $M_k$ gives

\begin{equation}
  \ddot{q}_k(t) + 2\zeta_k \omega_k\,\dot{q}_k(t) + \omega_k^2 q_k(t) = 0,
  \label{eq:modal_standard}
\end{equation}

where the modal damping ratio is

\begin{equation}
  \zeta_k = \frac{C_k}{2M_k\omega_k}.
\end{equation}

For the underdamped case ($0 < \zeta_k < 1$), the modal response is

\begin{equation}
  q_k(t)
  = e^{-\zeta_k \omega_k t}
    \left[
      A_k \cos(\omega_{dk} t)
      + B_k \sin(\omega_{dk} t)
    \right],
\end{equation}

where the damped natural frequency is
\begin{equation}
  \omega_{dk} = \omega_k \sqrt{1 - \zeta_k^2}.
\end{equation}

The modal constants $A_k$ and $B_k$ (not to be confused with the cross-sectional area $A$) are obtained by projection of the initial conditions

\begin{equation}
  w(x,0) = w_0(x), \qquad \dot{w}(x,0) = \dot{w}_0(x),
\end{equation}
onto the mode shapes:
\begin{align}
  A_k &= \frac{1}{M_k}\int_0^L \rho A\,\phi_k(x)\,w_0(x)\,dx, \\
  B_k &= \frac{1}{\omega_{dk}}
  \left[
    \frac{1}{M_k}\int_0^L \rho A\,\phi_k(x)\,\dot{w}_0(x)\,dx
    + \zeta_k \omega_k A_k
  \right].
\end{align}

The full displacement field is therefore

\begin{equation}
  w(x,t)
  = \sum_{k=1}^{\infty}
      \phi_k(x)\,
      e^{-\zeta_k \omega_k t}
      \left[
        A_k \cos(\omega_{dk} t)
        + B_k \sin(\omega_{dk} t)
      \right].
\end{equation}

This representation shows that the beam dynamics can be decomposed into a countable set of damped oscillatory modes, each governed by a second-order linear system. This modal structure provides a direct bridge to data-driven representations such as Dynamic Mode Decomposition (DMD), where analogous spatiotemporal modes and growth/decay rates are estimated directly from measurement data, provided the dynamics remain approximately linear within the observation window.

\subsection{Hankel Matrix Representation}

A fundamental tool in data-driven dynamical systems analysis is the Hankel matrix, which organizes a scalar or vector-valued time series into a structured matrix that encodes the temporal correlations of the signal. It is widely used in system identification, signal processing, and control theory. The Hankel matrix is characterized by the property that all elements along each anti-diagonal are constant.

Given a discrete time series $\{x_k\}_{k=1}^{N}$, the Hankel matrix is constructed as
\begin{equation}
\mathbf{H} =
\begin{bmatrix}
x_1 & x_2 & \cdots & x_q \\
x_2 & x_3 & \cdots & x_{q+1} \\
\vdots & \vdots & \ddots & \vdots \\
x_p & x_{p+1} & \cdots & x_{p+q-1}
\end{bmatrix},
\label{eq:hankel_matrix}
\end{equation}
where $p$ is the embedding dimension (number of delays), $q = N - p + 1$ is the number of columns (time-shifted snapshots), and $p + q - 1 = N$.

By incorporating delayed observations, the Hankel matrix implicitly transforms a one-dimensional signal into a higher-dimensional representation that captures the underlying temporal dependencies of the system, which facilitates the extraction of dominant spatio-temporal patterns, noise reduction, and low-rank approximations of system dynamics. 
The choice of embedding dimension $p$ (and delay, if applicable) is a critical design parameter. It must be sufficiently large to capture the essential dynamics while ensuring adequate data length for robust and well-conditioned analysis.

\subsection{Dynamic Mode Decomposition (DMD)}
Dynamic Mode Decomposition (DMD) is a data-driven framework for extracting spatio-temporal coherent structures from sequential data. It identifies a best-fit linear operator that advances the system state forward in time, enabling a spectral decomposition of the underlying dynamics. Despite its linear formulation, DMD provides a finite-dimensional approximation of the Koopman operator and is capable of capturing essential features of nonlinear systems.

First, two snapshot matrices are constructed from the set of snapshot pairs
\[
\left\{ \left( \mathbf{x}(t_i), \mathbf{x}(t_i') \right) \right\}_{i=1}^{m},
\]
where $t_i' = t_i + \Delta t$, and $\Delta t$ is the sampling interval chosen sufficiently small to resolve the highest frequencies in the dynamics. These snapshots are arranged into two data matrices,
\begin{equation}
\begin{aligned}
\mathbf{X} &=
\left[
\begin{array}{cccc}
\vert & \vert &        & \vert \\
\mathbf{x}(t_1) & \mathbf{x}(t_2) & \cdots & \mathbf{x}(t_m) \\
\vert & \vert &        & \vert
\end{array}
\right], \\
\\
\mathbf{X}' &=
\left[
\begin{array}{cccc}
\vert & \vert &        & \vert \\
\mathbf{x}(t_1') & \mathbf{x}(t_2') & \cdots & \mathbf{x}(t_m') \\
\vert & \vert &        & \vert
\end{array}
\right],
\end{aligned}
\label{eq:dmd_snapshots}
\end{equation}
where each column represents the system state at a given time instant.

The DMD algorithm seeks a best-fit linear operator $\mathbf{A}$ that establishes a linear dynamical system to advance snapshot measurements forward in time and relates these snapshot matrices as
\begin{equation}
\mathbf{X}' \approx \mathbf{A}\mathbf{X},
\label{eq:dmd_operator}
\end{equation}
For uniform sampling in time, this becomes
\begin{equation}
\mathbf{x}_{i+1} \approx \mathbf{A}\mathbf{x}_i.
\label{eq:dmd_dynamics}
\end{equation}

The optimal operator $\mathbf{A}$ is obtained in a least-squares sense as
\begin{equation}
\mathbf{A} = \arg\min_{\mathbf{A}} \left\| \mathbf{X}' - \mathbf{A}\mathbf{X} \right\|_F
= \mathbf{X}' \mathbf{X}^\dagger,
\label{eq:dmd_A}
\end{equation}
where $\mathbf{X}^{\dagger}$ denotes the Moore--Penrose pseudoinverse and $\|\cdot\|_F$ denotes the Frobenius norm.

For high-dimensional systems, direct computation of $\mathbf{A}$ is computationally expensive. Therefore, a reduced-order approximation is constructed using the singular value decomposition (SVD) of $\mathbf{X}$,
\begin{equation}
\mathbf{X} = \mathbf{U} \boldsymbol{\Sigma} \mathbf{V}^*.
\label{eq:dmd_svd}
\end{equation}

Since the snapshot matrix typically has low intrinsic rank, only the leading $r$ singular values and corresponding singular vectors are retained, yielding the truncated decomposition
\begin{equation}
\mathbf{X} \approx \mathbf{U}_r \boldsymbol{\Sigma}_r \mathbf{V}_r^*,
\label{eq:dmd_truncated_svd}
\end{equation}
where $r \leq \min(n,m)$ is the chosen truncation rank. The selection of $r$ is a critical step, as it determines the balance between capturing dominant dynamics and filtering noise.

Projecting the dynamics onto this reduced subspace yields the low-dimensional operator
\begin{equation}
\tilde{\mathbf{A}} = \mathbf{U}_r^* \mathbf{X}' \mathbf{V}_r \boldsymbol{\Sigma}_r^{-1},
\label{eq:dmd_Atilde}
\end{equation}
which has dimension $r \times r$ and shares the same nonzero eigenvalues as the full operator $\mathbf{A}$.

The spectral decomposition of $\tilde{\mathbf{A}}$ is given by
\begin{equation}
\tilde{\mathbf{A}} \mathbf{W} = \mathbf{W} \boldsymbol{\Lambda},
\label{eq:dmd_eig}
\end{equation}
where $\boldsymbol{\Lambda}$ contains the DMD eigenvalues and $\mathbf{W}$ contains the corresponding eigenvectors.

The DMD modes are reconstructed as
\begin{equation}
\boldsymbol{\Phi} = \mathbf{X}' \mathbf{V}_r \boldsymbol{\Sigma}_r^{-1} \mathbf{W}.
\label{eq:dmd_modes}
\end{equation}

The columns of $\boldsymbol{\Phi}$ are referred to as DMD modes, which represent spatially coherent structures that evolve in time with fixed oscillation frequencies and growth or decay rates. The eigenvalues $\lambda_k$ contained in $\boldsymbol{\Lambda}$ are the DMD eigenvalues and characterize the temporal behavior of each mode, encoding both oscillatory dynamics and exponential growth or decay.

\subsection{Damage Detection Algorithms}

Vibration-based damage detection relies on the principle that structural damage reduces local stiffness, which in turn perturbs the mode shapes and amplifies their spatial derivatives near the damage location~\cite{Pandey1991}. In this study, a novel damage identification framework is developed using Dynamic Mode Decomposition (DMD) modes as a substitute for classical 
experimental mode shapes. Spatial features are extracted from each DMD mode profile and compared against a healthy reference state to identify and localize structural changes.

\paragraph{Spatial Feature Extraction}

From each DMD mode profile $\phi_k(x)$, four spatial feature vectors are extracted, corresponding to progressively higher-order spatial derivatives:
\begin{align}
  \text{MS:}   \quad & \mathbf{x}_k^{(1)} = \phi_k(x), 
                       \label{eq:feat_MS}   \\
  \text{MSS:}  \quad & \mathbf{x}_k^{(2)} = \frac{d\phi_k}{dx}, 
                       \label{eq:feat_MSS}  \\
  \text{MSC:}  \quad & \mathbf{x}_k^{(3)} = \frac{d^2\phi_k}{dx^2}, 
                       \label{eq:feat_MSC}  \\
  \text{MSCS:} \quad & \mathbf{x}_k^{(4)} = \left(\frac{d^2\phi_k}{dx^2}
                       \right)^{\!2},
                       \label{eq:feat_MSCS}
\end{align}

where MS, MSS, MSC, and MSCS denote the mode shape, mode shape slope, mode shape curvature, and mode shape curvature square, respectively. The physical motivation for these features is rooted in Euler-Bernoulli beam theory: the curvature $d^2\phi/dx^2$ is directly proportional to the 
local bending strain via $\phi'' = \mathcal{M}/(EI)$, so a local reduction in flexural rigidity $EI$ caused by damage produces a localized anomaly in the curvature and its square~\cite{Pandey1991, Wahab1999}.

In the discrete setting, spatial derivatives are approximated using central differences. For measurement locations uniformly spaced by $h$, the first and second derivatives at interior node $i$ are
\begin{align}
  {v_i'}^{(k)}  &= \frac{v_{i+1}^{(k)} - v_{i-1}^{(k)}}{2h}, 
                   \label{eq:slope}    \\
  {v_i''}^{(k)} &= \frac{v_{i+1}^{(k)} - 2v_i^{(k)} + v_{i-1}^{(k)}}{h^2},
                   \label{eq:curvature}
\end{align}
where $v_i^{(k)}$ denotes the $k$-th mode shape component at spatial location $i$. Prior to feature extraction, all mode shapes must be consistently normalized (e.g.\ mass-normalized or unit-norm) to ensure a physically meaningful comparison between the damaged and reference states.

For each feature type, the feature vectors from the $M$ selected DMD modes are assembled into a feature matrix
\begin{equation}
  \mathbf{X} = \bigl[\mathbf{x}_1,\, \mathbf{x}_2,\, \dots,\, 
  \mathbf{x}_M\bigr] \in \mathbb{R}^{n \times M},
  \label{eq:feature_matrix}
\end{equation}
where $n$ is the number of spatial measurement locations and $M$ is the 
number of selected modes.

\paragraph{Baseline Reference Construction}

The healthy (undamaged) state is used to construct a baseline reference for each feature type. The reference matrix $\mathbf{P}$ is defined as a variance-weighted outer-product accumulation over the $M$ healthy modes,

\begin{equation}
  \mathbf{P} = \frac{C}{M} \sum_{k=1}^{M} 
  \frac{\mathbf{x}_k^{\mathrm{healthy}}
  \left(\mathbf{x}_k^{\mathrm{healthy}}\right)^{\!\top}}{\sigma_k^2},
  \label{eq:baseline}
\end{equation}

where $C$ is a positive scaling constant and
\begin{equation}
  \sigma_k^2 = \mathrm{Var}\!\left(\mathbf{x}_k^{\mathrm{healthy}}\right)
  \label{eq:variance}
\end{equation}

is the spatial variance of the $k$-th healthy feature vector, which normalizes the contribution of each mode and prevents modes with large dynamic range from dominating the reference. To ensure numerical invertibility, a regularization term is added,

\begin{equation}
  \mathbf{P} \leftarrow \mathbf{P} + \epsilon\,\mathbf{I},
  \label{eq:regularisation}
\end{equation}
where $\epsilon \ll 1$ is a small positive scalar and $\mathbf{I}$ is the identity matrix.

\paragraph{Damage Index Formulation}

For a given mode $k$, the deviation of the current (potentially damaged) feature vector from the healthy reference is

\begin{equation}
  \mathbf{d}_k = \mathbf{x}_k^{\mathrm{current}} - 
  \mathbf{x}_k^{\mathrm{healthy}}.
  \label{eq:deviation}
\end{equation}
A mode-wise damage index is then defined as the Mahalanobis-type quadratic form

\begin{equation}
  Q_k = \mathbf{d}_k^{\top} \left(2\mathbf{P}\right)^{-1} \mathbf{d}_k, 
  \qquad k = 1, 2, \dots, M,
  \label{eq:Qk}
\end{equation}

which weights the deviation by the inverse of the reference covariance structure, thereby accounting for the natural variability in each spatial feature. A global damage index is obtained by averaging over the selected modes,

\begin{equation}
  Q = \frac{1}{M} \sum_{k=1}^{M} Q_k,
  \label{eq:Q_global}
\end{equation}

with larger values of $Q$ indicating a greater departure from the healthy baseline and hence a higher likelihood of structural damage. At the spatial level, pointwise damage indices analogous to those in Eqs.~\eqref{eq:feat_MS}--\eqref{eq:feat_MSCS} can be recovered by inspecting the individual components of $\mathbf{d}_k$, allowing spatial localization of the damage.

\begin{figure}[H]
\centering
\begin{picture}(400,340)
    \put(40,185){\includegraphics[width=0.75\columnwidth]{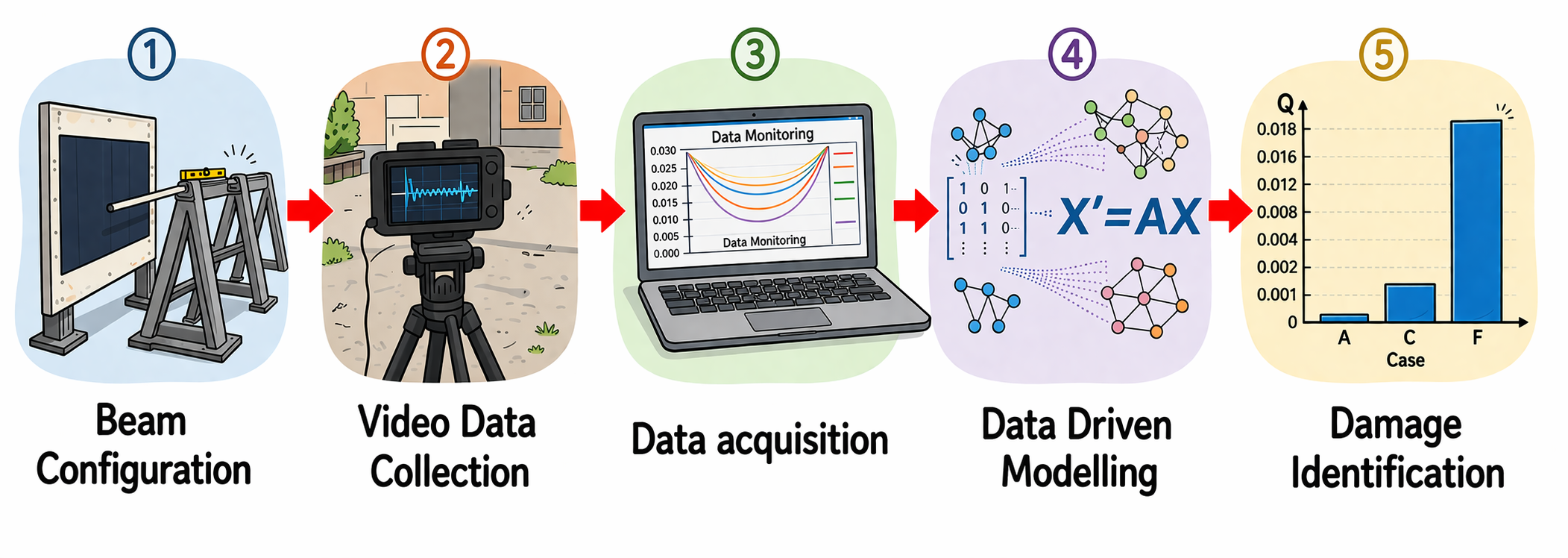}}
    \put(40,20){\includegraphics[width=0.75\columnwidth]{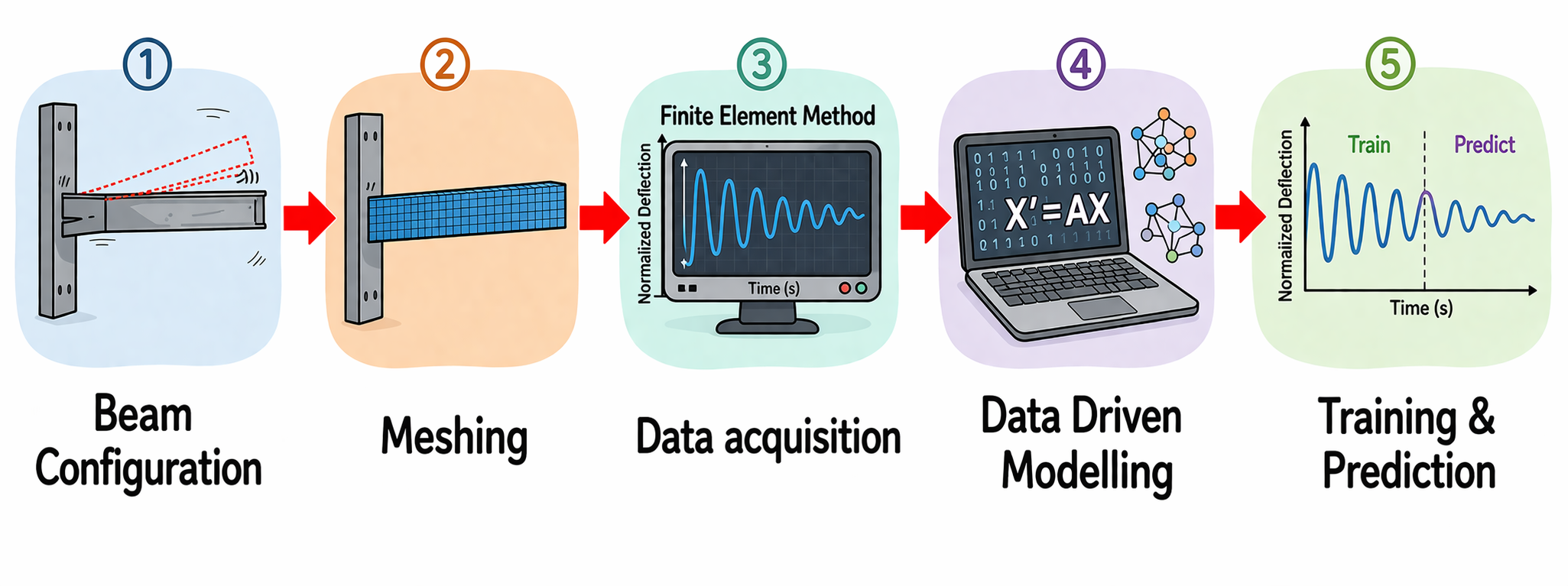}}

    \put(45,320){\large \textbf{(a)}}
    \put(45,155){\large \textbf{(b)}}
\end{picture}

\vspace{0.1cm}
\caption{Overview of the proposed damage identification and structural response prediction from video data: (a) Experimental workflow; (b) numerical workflow.}
\label{fig:workflow_charts}
\end{figure}

\section{Data Generation}\label{data_generation}

This study adopts a twofold approach to investigate vibration-based damage detection in a cantilever beam, combining a finite element simulation study with a companion experimental study. Both approaches consider the same structural configuration and damage scenarios considered in this study.

Simulation data are generated through transient finite element analyses carried out in ANSYS. The structural model consists of a polypropylene cantilever beam with a total length of 0.8~m and a square cross-section of $25.4 \times 25.4$~mm, subject to fixed-free boundary conditions. A mesh element size of
$2 \times 10^{-2}$ m is used for all simulations. Free vibration is initiated by prescribing an initial tip displacement, after which the transient displacement field is extracted at each time step for the full beam length. One healthy (baseline) state and two damaged configurations are analysed, with the damage scenarios summarised in Table~\ref{tab:damage_cases_acf}. For the damaged configurations, each crack is assigned a thickness of 1 mm. The corresponding mesh statistics are 1221 nodes
and 160 elements for the healthy case, 1366 nodes and 182 elements for Damage Case 1, and 1651 nodes and 228 elements for
Damage Case 2. The extracted displacement fields serve as the input to the DMD-based identification framework described in Section~\ref{sec:method}.

The experimental data used in this study are sourced from Garrido et al.~\cite{Garrido2024}, in which a polypropylene cantilever beam of total length 1~m and square cross-section $25.4 \times 25.4$~mm was tested under free vibration. The beam was clamped at the right end, yielding an effective free span of 0.8~m. In the undamaged condition, the fundamental natural period (FNP) was measured as 0.13~s. Free vibration was induced by applying an initial quasi-static tip 
displacement of approximately 4--5~cm, which was then released. The subsequent transverse response was recorded using a Chronos 2.1-HD high-speed camera at a frame rate of 1000~fps with a full exposure time of 1/1000~s, over a total duration of 2.7~s (approximately 20 fundamental natural periods). The camera resolution was set to $1280 \times 1024$~pixels, of which a region of interest (ROI) comprising 1200 pixels along the beam span was retained for analysis. 
The optical setup yielded a spatial resolution of approximately 0.6 mm/pixel. Damage was introduced in the form of transverse rectangular notches on the upper surface of the beam, machined using a steel saw. The notches approximate Mode~I (opening) cracks with widths in the range 0.5-1 mm, sufficiently narrow to be treated as infinitesimal cracks while 
preventing crack closure (breathing) during vibration. Crack locations were placed at 100 mm, 200~mm, and 300~mm from the fixed end, and crack depths were measured with a Vernier caliper prior to testing. Three representative damage scenarios are selected from the full dataset for comparison with the simulation study, and are summarised in Table~\ref{tab:damage_cases_acf}.

\begin{table}[H]
\caption[Damage cases A, C, and F]{Depth and location of damage from the free end for selected cases.}
\label{tab:damage_cases_acf}
\centering
\begin{tabular}{>{\centering\arraybackslash}p{2.2cm}ccc}
\toprule
\textbf{Case Name} & \multicolumn{3}{c}{\textbf{Depth and Location of Damage}} \\
                   & \multicolumn{3}{c}{\textbf{from Free End}} \\
\cmidrule(lr){2-4}
                   & at 0.5\,m & at 0.6\,m & at 0.7\,m \\
\midrule
Healthy Case  & --    & --    & --    \\
Damage Case 1 & --    & --    & 10\,mm \\
Damage Case 2 & 5\,mm & 5\,mm & 13\,mm \\
\bottomrule
\end{tabular}
\end{table}

\section{Results and Discussion}

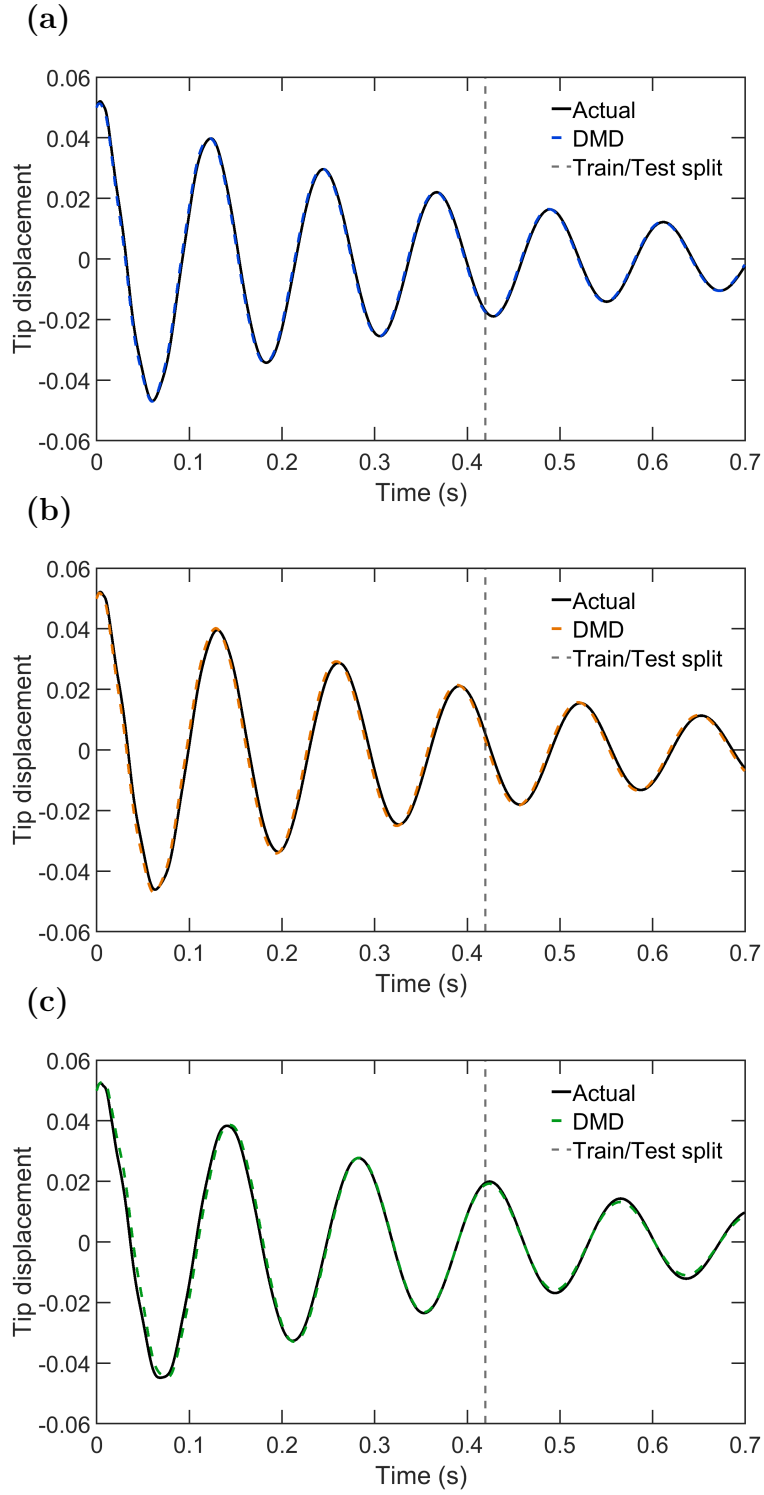
\begin{figure}

\begin{picture}(400,590)
    \put(90,380){\includegraphics[width=0.6\columnwidth]{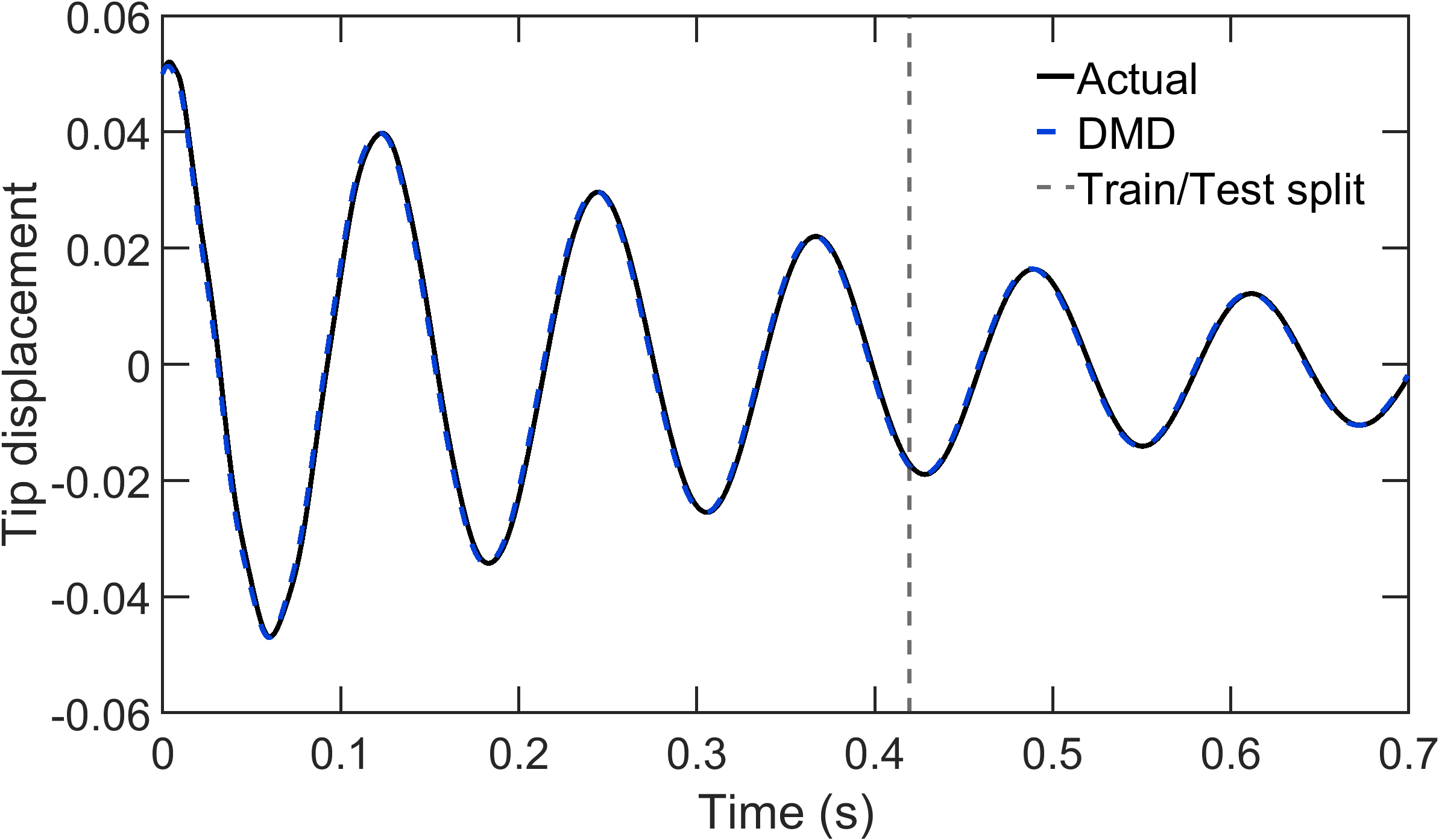}}
    \put(90,195){\includegraphics[width=0.6\columnwidth]{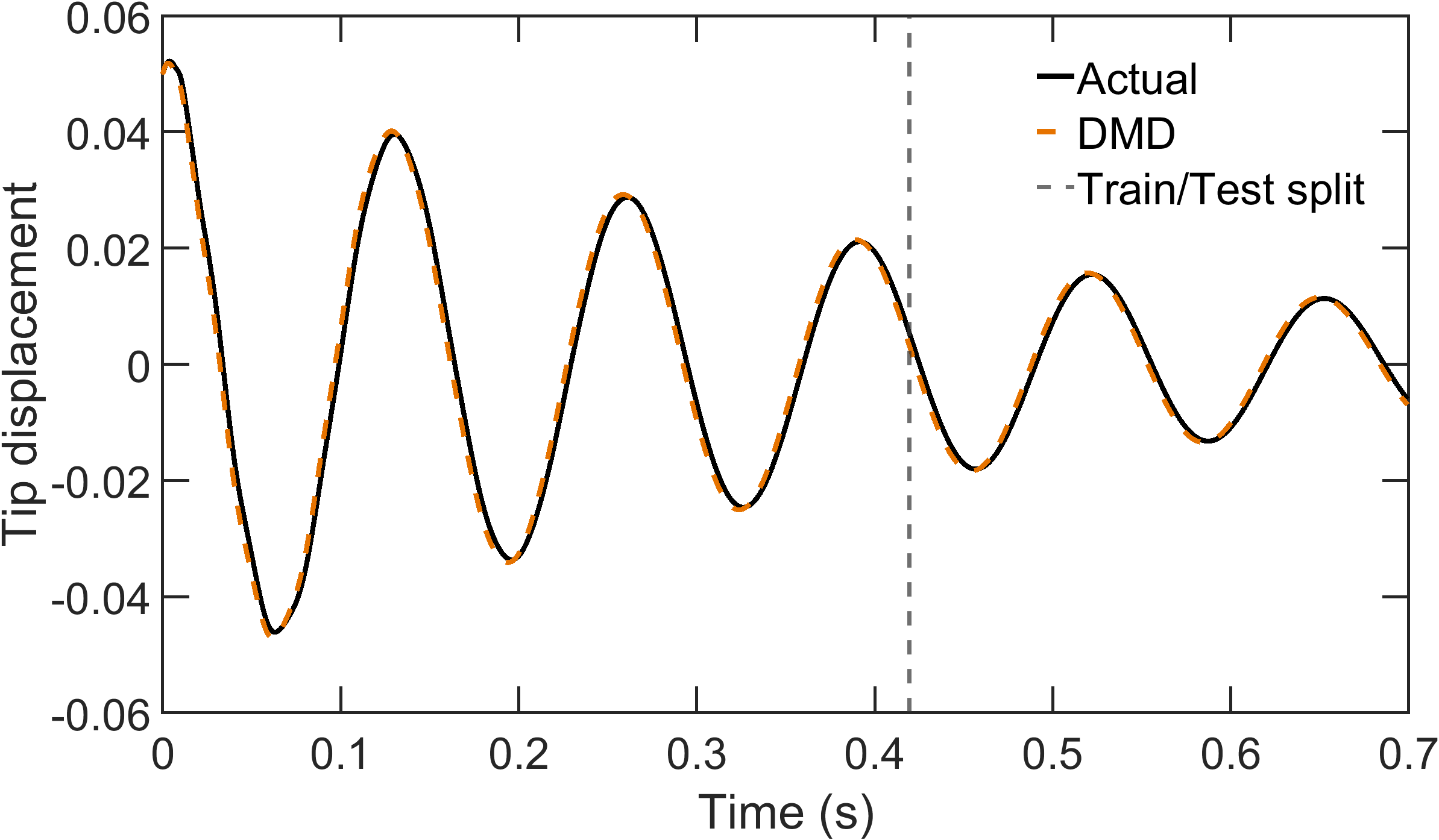}}
    \put(90,10){\includegraphics[width=0.6\columnwidth]{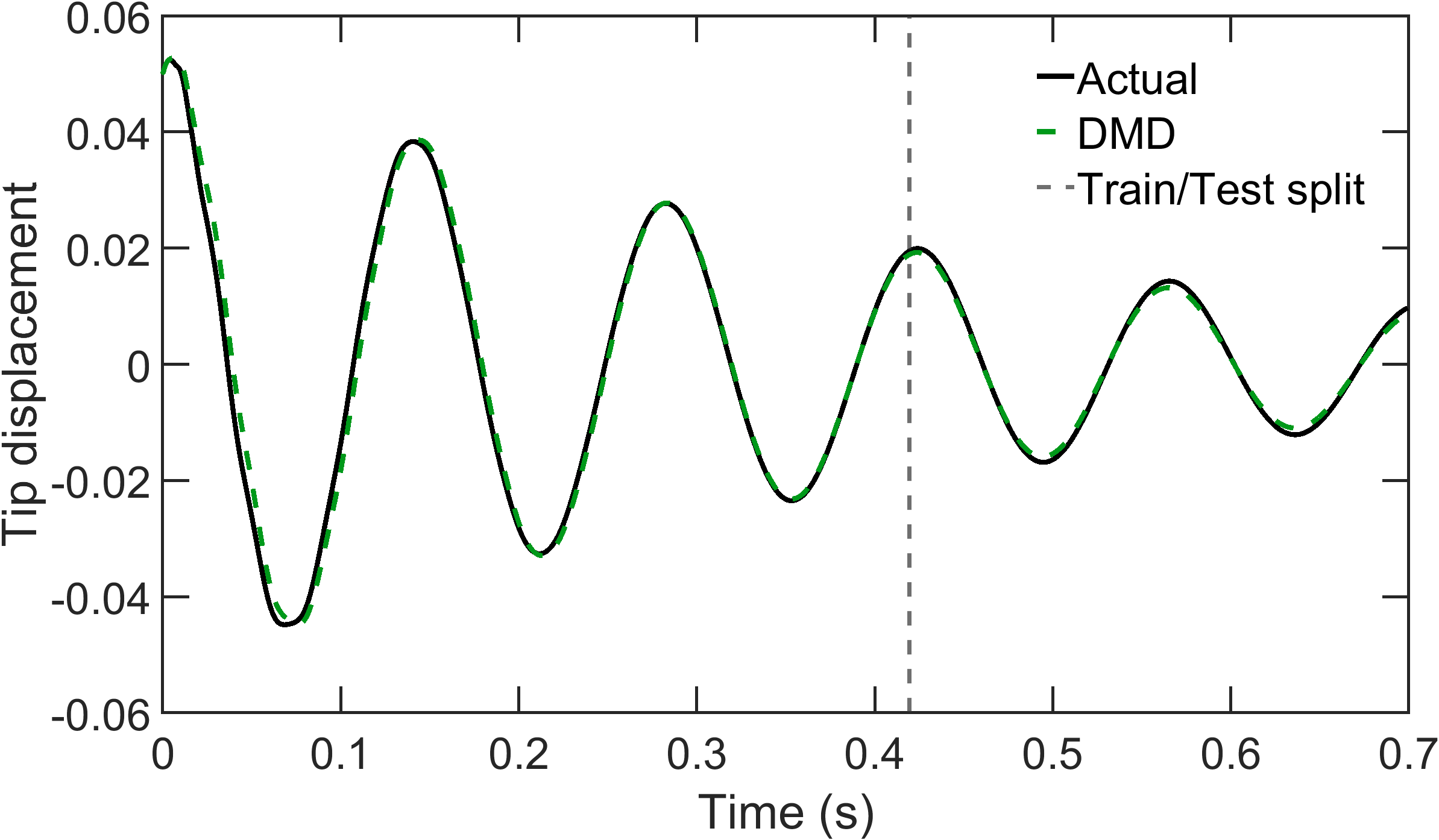}}

    \put(95,560){\large \textbf{(a)}}
    \put(95,375){\large \textbf{(b)}}
    \put(95,190){\large \textbf{(c)}}
\end{picture}

\vspace{0.1cm}\caption{ Comparison of the DMD reconstruction and prediction for the tip displacement and the actual finite-element simulation: (a) Healthy; (b) Damage Case 1; and (c) Damage Case 2.}
\label{fig:dmd_reconstruction_cases}
\end{figure}

In this section, the findings from both the simulation-based and experimental studies are presented. The simulation study focuses on the application of DMD algorithm for reconstructing and predicting the vibration response in healthy and damaged configurations, followed by a comparative analysis of the extracted DMD modes to examine variations in the underlying structural dynamics. Subsequently, the experimental study presents the DMD eigenvalue characteristics across different structural states. In addition, damage detection is performed using the damage index $Q$ (Eq. \ref{eq:Q_global}), enabling a thorough assessment of the sensitivity of different DMD-based features such as MS, MSS, MSC, and MSCS.

The reconstruction and predictive performance of Dynamic Mode Decomposition (DMD) for the simulated cases are illustrated in Figure~\ref{fig:dmd_reconstruction_cases}, where the tip displacement obtained from the high-fidelity model is compared with the DMD-based estimates for both healthy and damaged configurations. To better capture the underlying system dynamics, delay embedding was employed, effectively augmenting the state representation with time-lagged measurements and implicitly incorporating velocity information. The DMD model was identified using the first $60\%$ of the available time samples, while the remaining data were reserved for out-of-sample prediction. As observed, DMD provides an accurate reconstruction of the response within the training interval. More importantly, the model exhibits strong predictive capability beyond the training window, maintaining consistency with the true system dynamics in the extrapolation regime.

The reconstruction and prediction accuracy are further quantified by the root-mean-square error (RMSE) for all cases. The healthy configuration yields the lowest error ($1.12 \times 10^{-3}$), whereas slightly higher errors are observed for the damaged configurations ($2.13 \times 10^{-3}$ for Damage Case 1 and $2.08 \times 10^{-3}$ for Damage Case 2). This increase in error suggests that structural damage introduces additional dynamical complexity—such as changes in modal properties or increased nonlinearity—which is more challenging to capture using a linear DMD framework. Nevertheless, the error levels remain consistently low across all cases, demonstrating that DMD provides a robust representation of the dominant system dynamics. Furthermore, the accurate preservation of both phase and amplitude in the prediction regime indicates that the principal modal contributions are effectively identified and propagated by the DMD model.

\begin{figure}[htb]

\begin{picture}(400,360)
    \put(-10,180){\includegraphics[width=0.5\columnwidth]{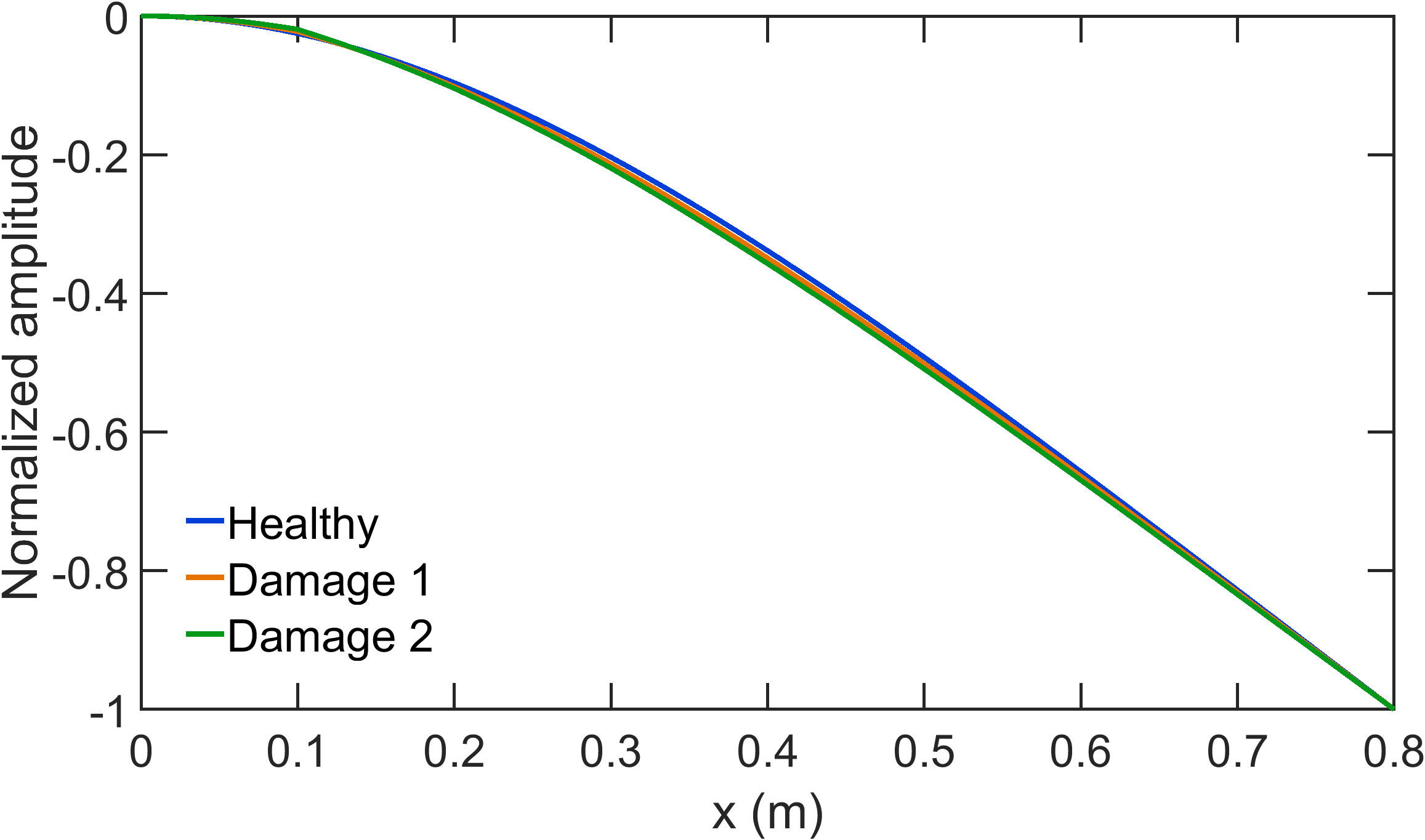}}
    \put(230,180){\includegraphics[width=0.5\columnwidth]{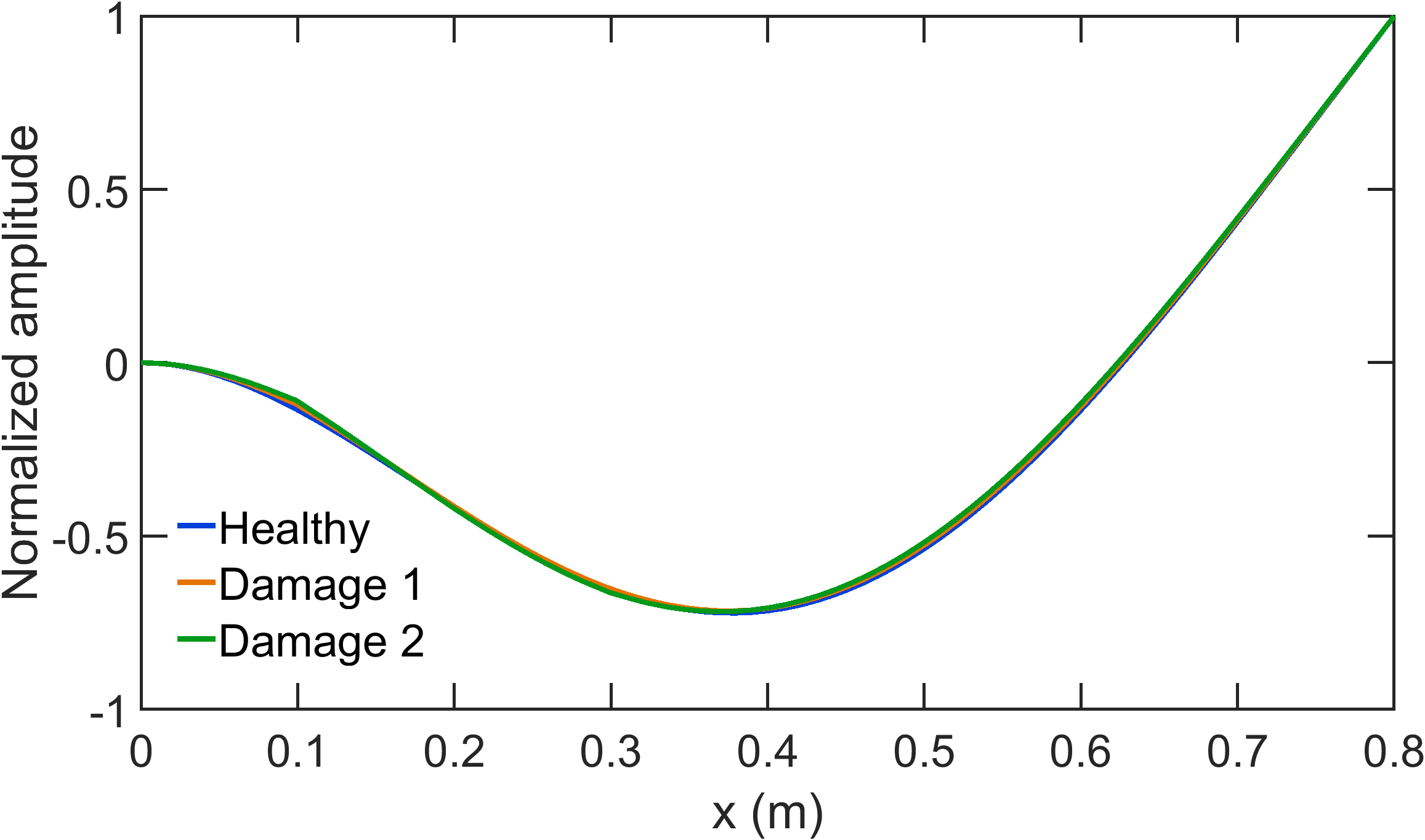}}
    \put(-10,0){\includegraphics[width=0.5\columnwidth]{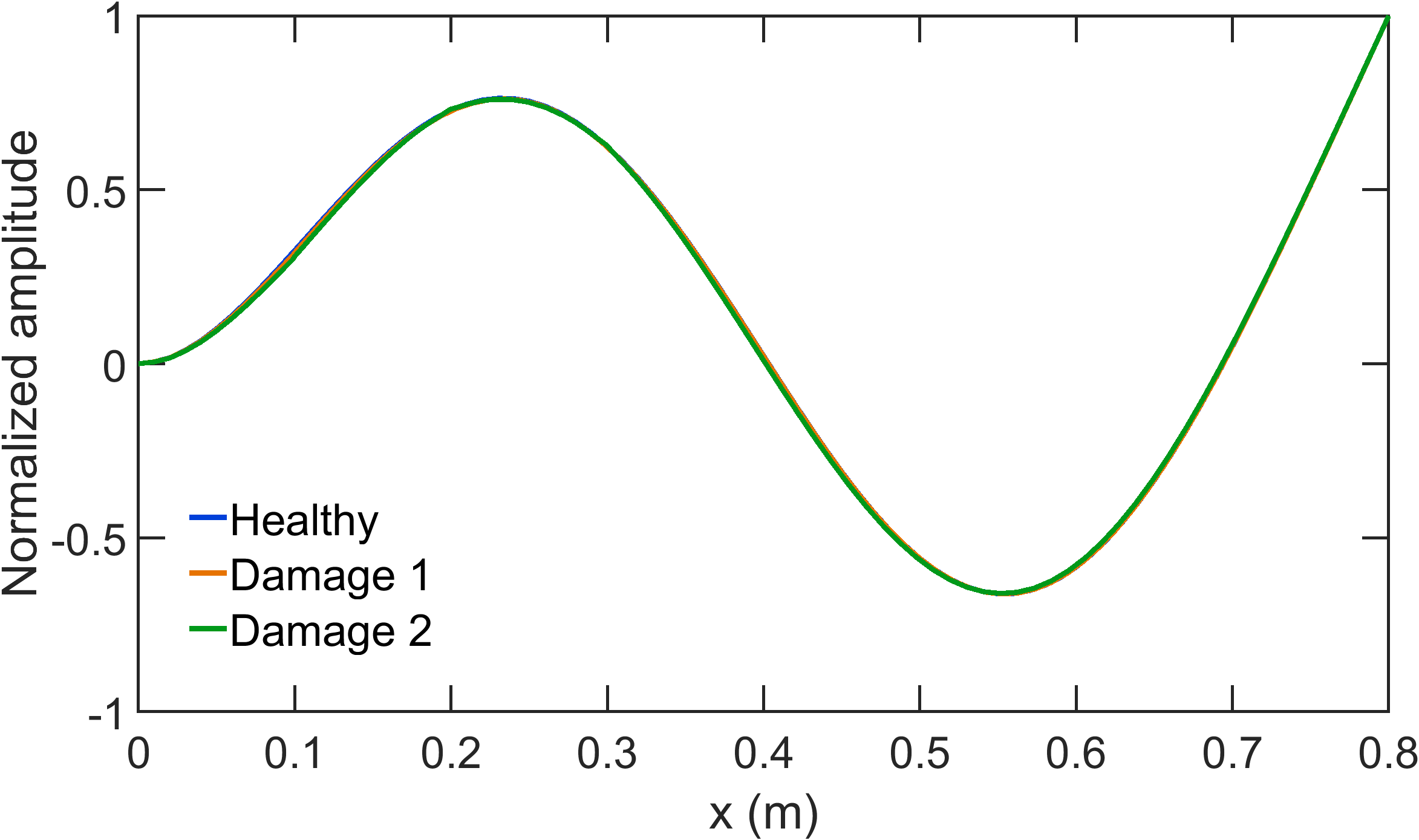}}
    \put(230,0){\includegraphics[width=0.5\columnwidth]{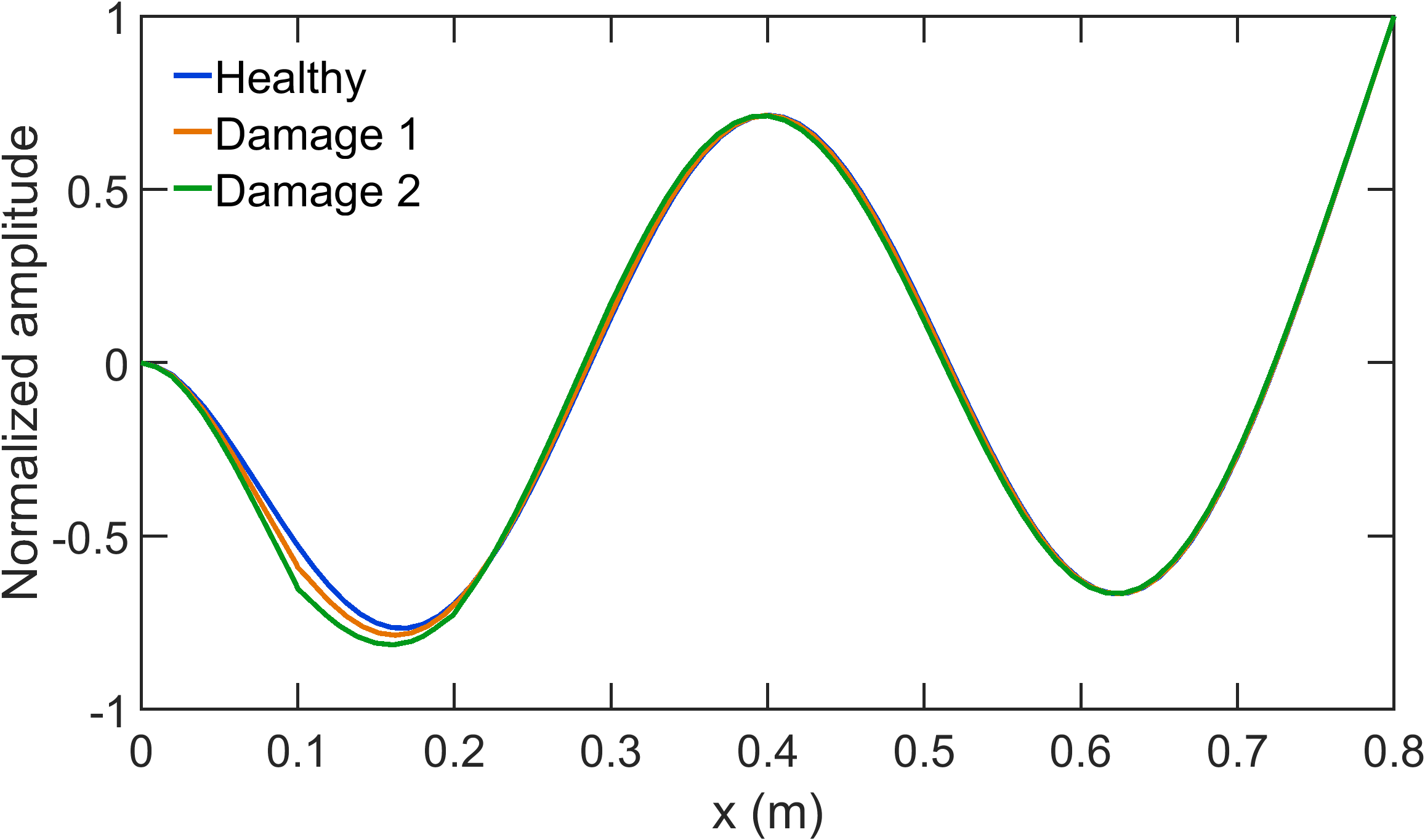}}

    \put(20,335){\large \textbf{(a)}}
    \put(280,335){\large \textbf{(b)}}
    \put(20,155){\large \textbf{(c)}}
    \put(280,155){\large \textbf{(d)}}
\end{picture}

\vspace{0.5cm}\caption{Mode shapes obtained from the displacement field of the simulation study using DMD algorithm for different structural states: (a) first mode; (b) second mode; (c) third mode; and (d) fourth mode.}
\label{fig:simulation_modes}
\end{figure}

The DMD modes extracted for the healthy and damaged configurations are presented in Figure~\ref{fig:simulation_modes}, providing insight into variations in the underlying structural dynamics. The dominant spatial modes are normalized to enable a consistent comparison across different cases. The healthy configuration exhibits smooth and physically consistent mode shapes, characteristic of an undamaged cantilever beam with uniform stiffness distribution. In contrast, the damaged configurations display subtle but systematic deviations in the spatial structure of the modes, indicating alterations in the dynamic characteristics of the system.

\begin{figure}[htb]

\begin{picture}(400,360)
    \put(-10,180){\includegraphics[width=0.5\columnwidth]{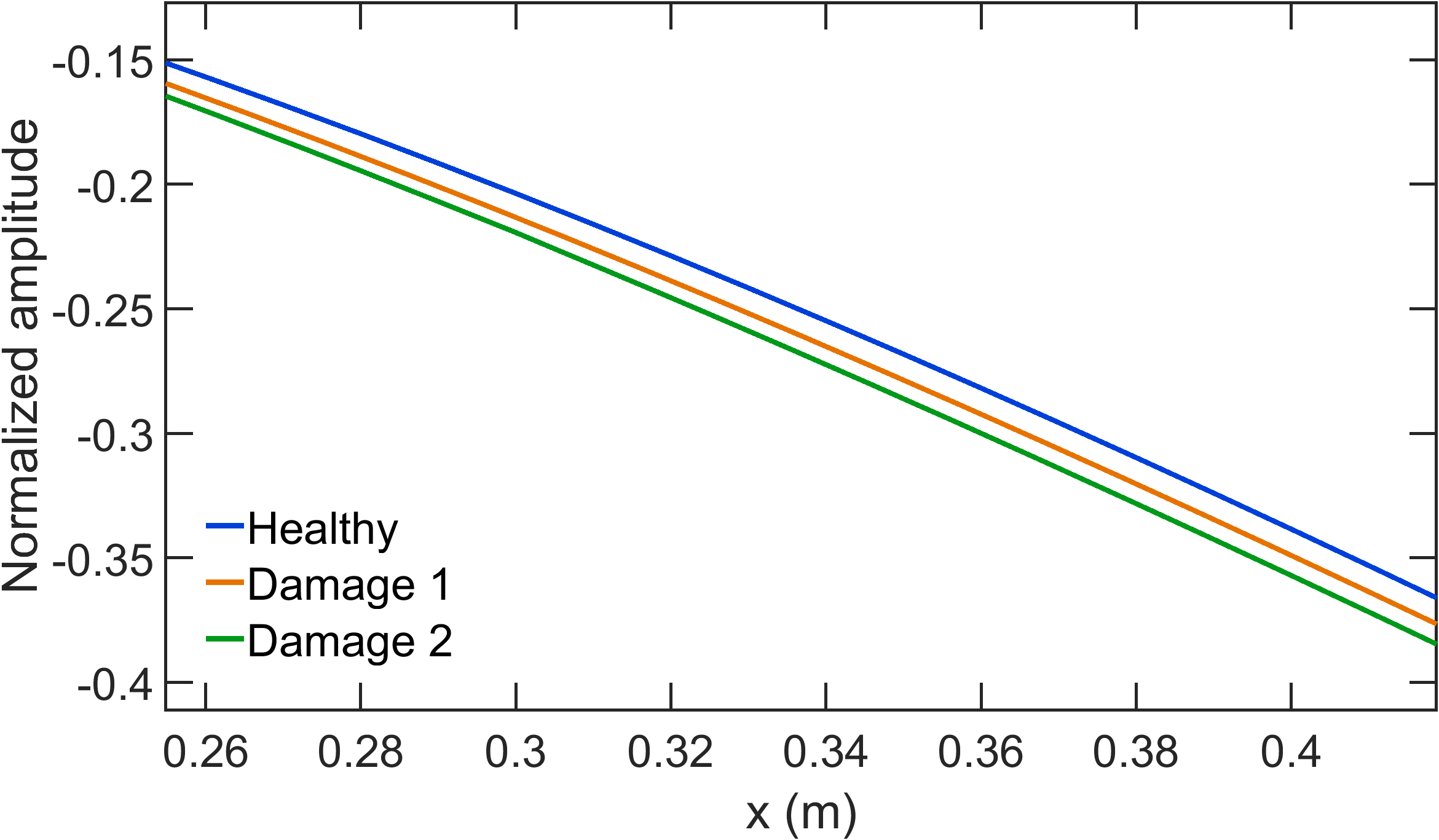}}
    \put(230,180){\includegraphics[width=0.5\columnwidth]{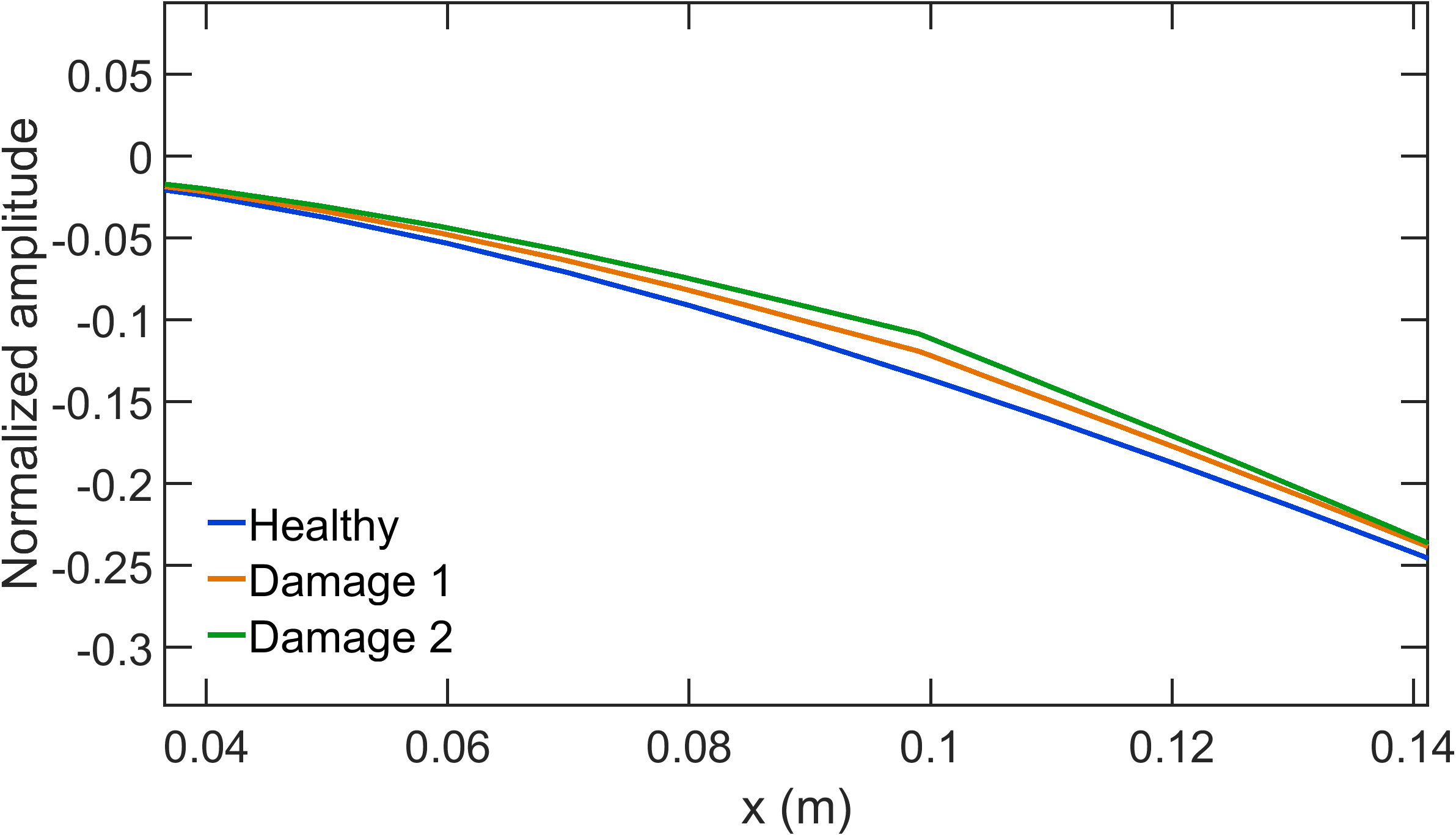}}
    \put(-10,0){\includegraphics[width=0.5\columnwidth]{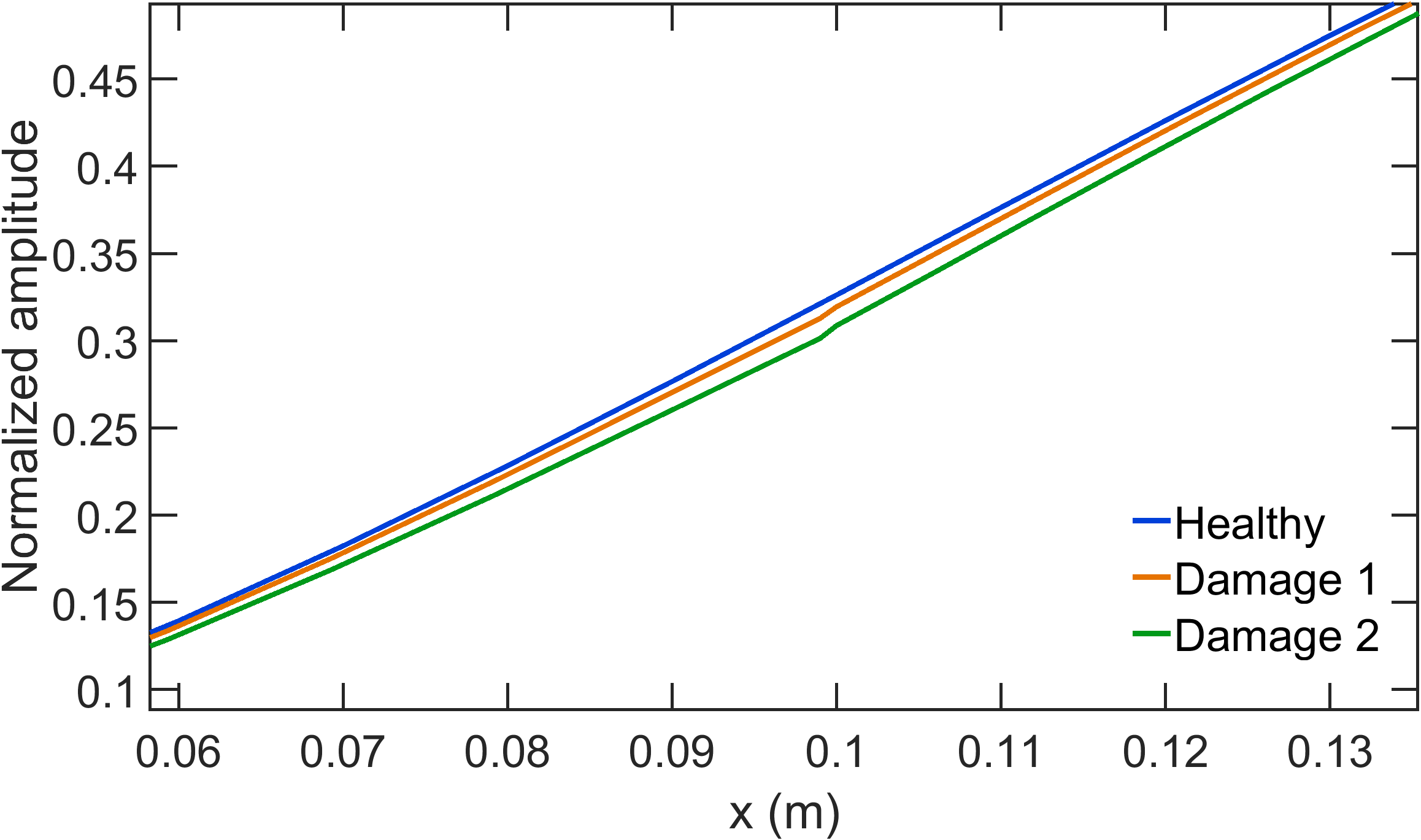}}
    \put(230,0){\includegraphics[width=0.5\columnwidth]{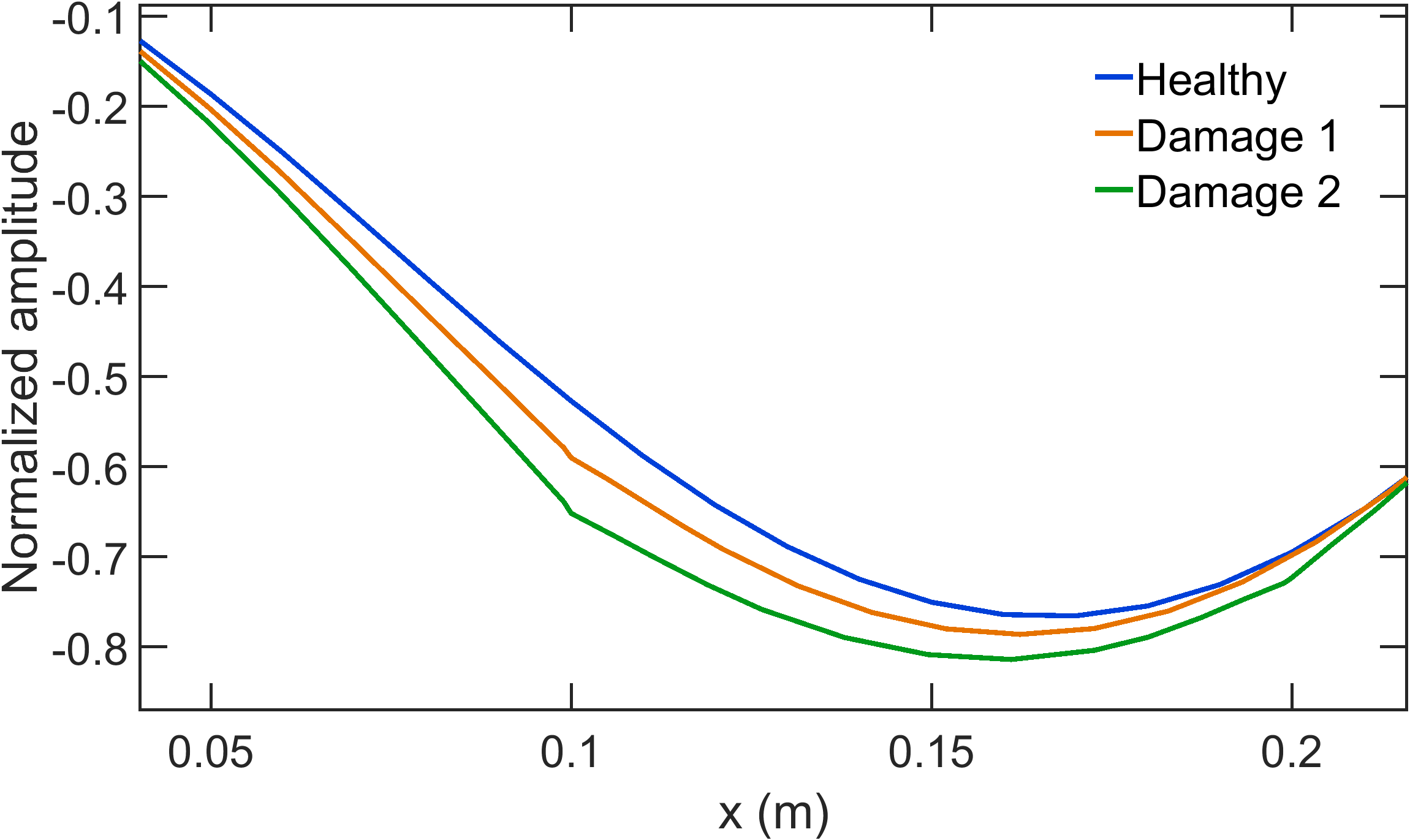}}

    \put(20,335){\large \textbf{(a)}}
    \put(280,335){\large \textbf{(b)}}
    \put(20,155){\large \textbf{(c)}}
    \put(280,155){\large \textbf{(d)}}
\end{picture}

\vspace{0.5cm}\caption{Zoomed view of simulation mode shapes: (a) first mode; (b) second mode; (c) third  mode; and (d) fourth  mode.}
\label{fig:simulation_modes_zoomed}
\end{figure}

These differences become more evident in the zoomed views shown in Figure~\ref{fig:simulation_modes_zoomed}, where localized deviations in modal amplitude and curvature can be observed. Such variations are indicative of changes in the effective stiffness distribution and, consequently, shifts in the modal properties induced by damage. While the global shape of the modes remains largely preserved, the localized distortions reflect modifications in the spatial energy distribution of the system. This behavior highlights the sensitivity of DMD-derived modes to structural perturbations and demonstrates their capability to capture damage-induced changes in the dynamic response, even when the overall modal structure appears qualitatively similar.

\begin{figure}[!t]
\begin{picture}(400,230)
    \put(-10,20){\includegraphics[width=0.5\columnwidth]{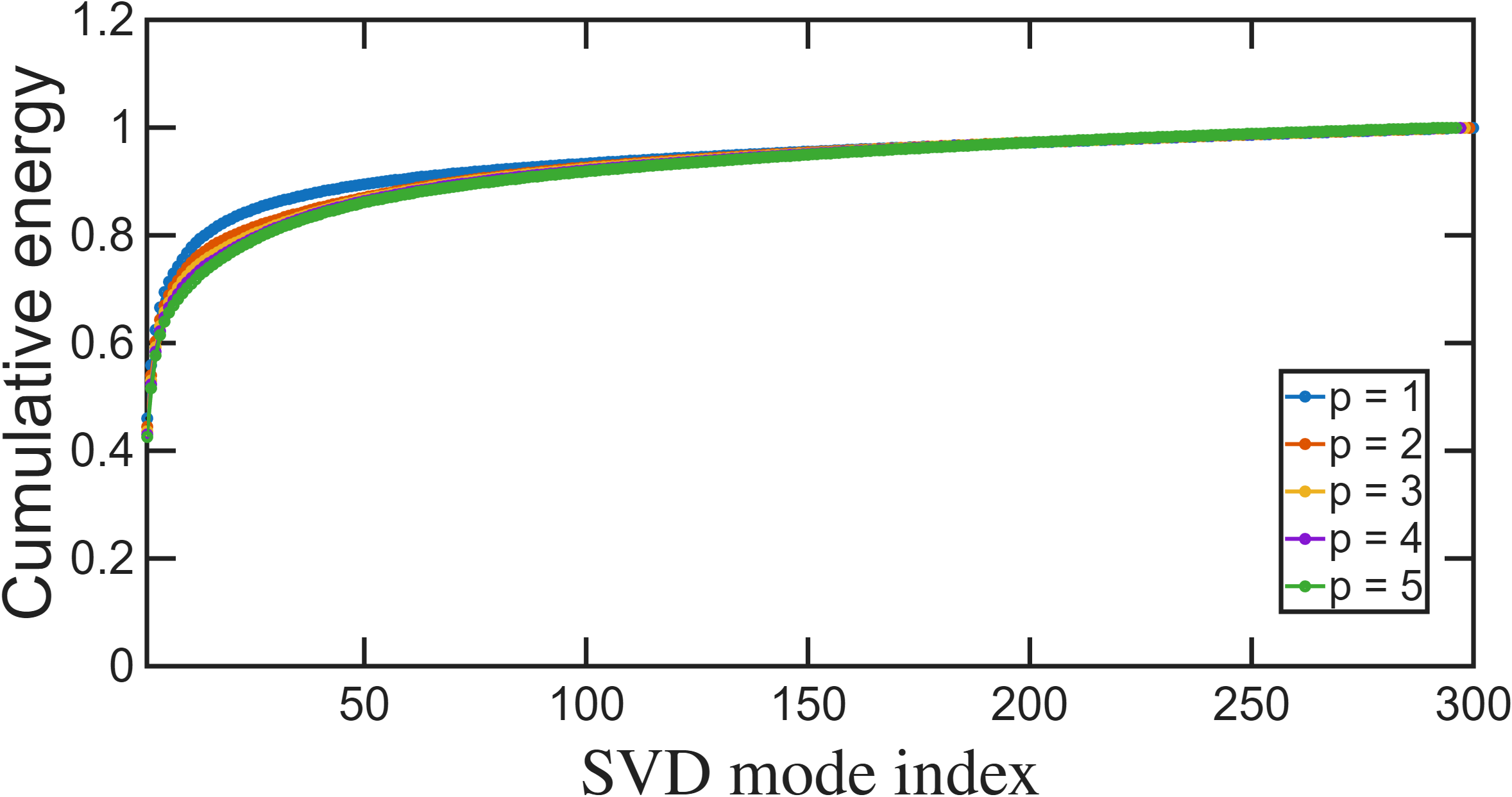}}
    \put(230,20){\includegraphics[width=0.5\columnwidth]{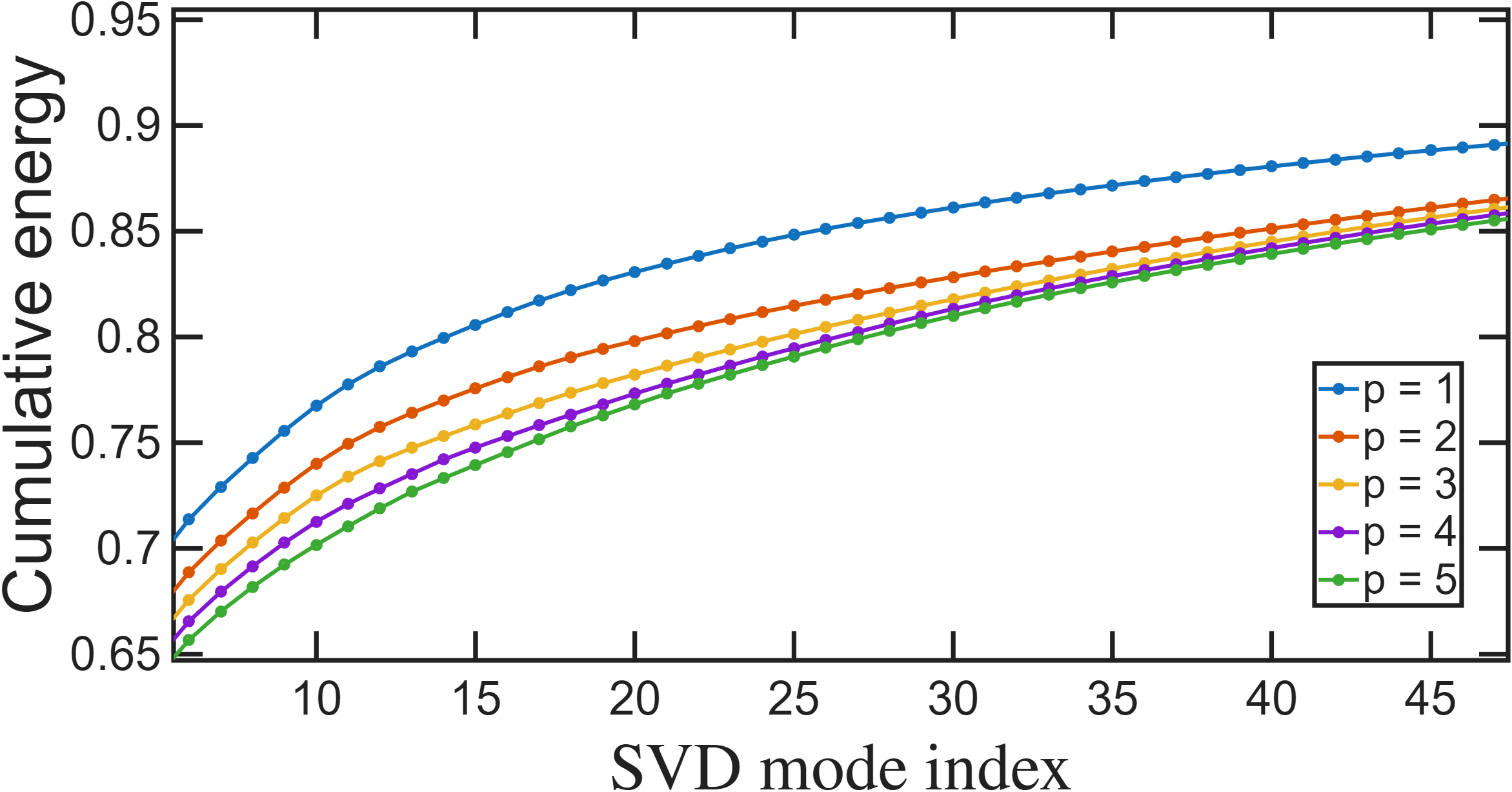}}

    \put(20,150){\large \textbf{(a)}}
    \put(280,150){\large \textbf{(b)}}
\end{picture}

\vspace{-0.3cm}
\caption{Cumulative energy over Hankel embedding: (a) full view; and (b) zoomed view.}
\label{fig:cumulative_energy_combined}
\end{figure}

For the experimental video data, delay embedding can be employed to construct a Hankel matrix in order to enrich the state representation and better capture the underlying system dynamics. To guide the selection of an appropriate embedding dimension, a cumulative energy criterion based on singular value decomposition (SVD) was used. The cumulative energy distributions for different Hankel embedding dimensions are presented in Figure~\ref{fig:cumulative_energy_combined}. As observed, increasing the embedding dimension does not significantly alter the overall trend of energy accumulation, and all cases exhibit a similar decay behavior in the singular value spectrum. This observation suggests that, for the present dataset, delay embedding does not substantially increase the amount of dominant energy captured by the leading modes, but primarily redistributes the information across a higher-dimensional representation. Although the cumulative energy increases gradually over a wide range of truncation ranks, a rank in the range of approximately $150$--$250$ is sufficient to capture the majority of the system’s energetic content. Based on this trade-off between model fidelity and complexity, a truncation rank of $150$ without delay embedding was selected for subsequent analysis. This choice preserves the dominant dynamics while avoiding unnecessary model expansion, and all experimental DMD results reported herein are based on this configuration.

\begin{figure}[!b]

\begin{picture}(400,250)
    \put(-10,20){\includegraphics[width=0.48\columnwidth]{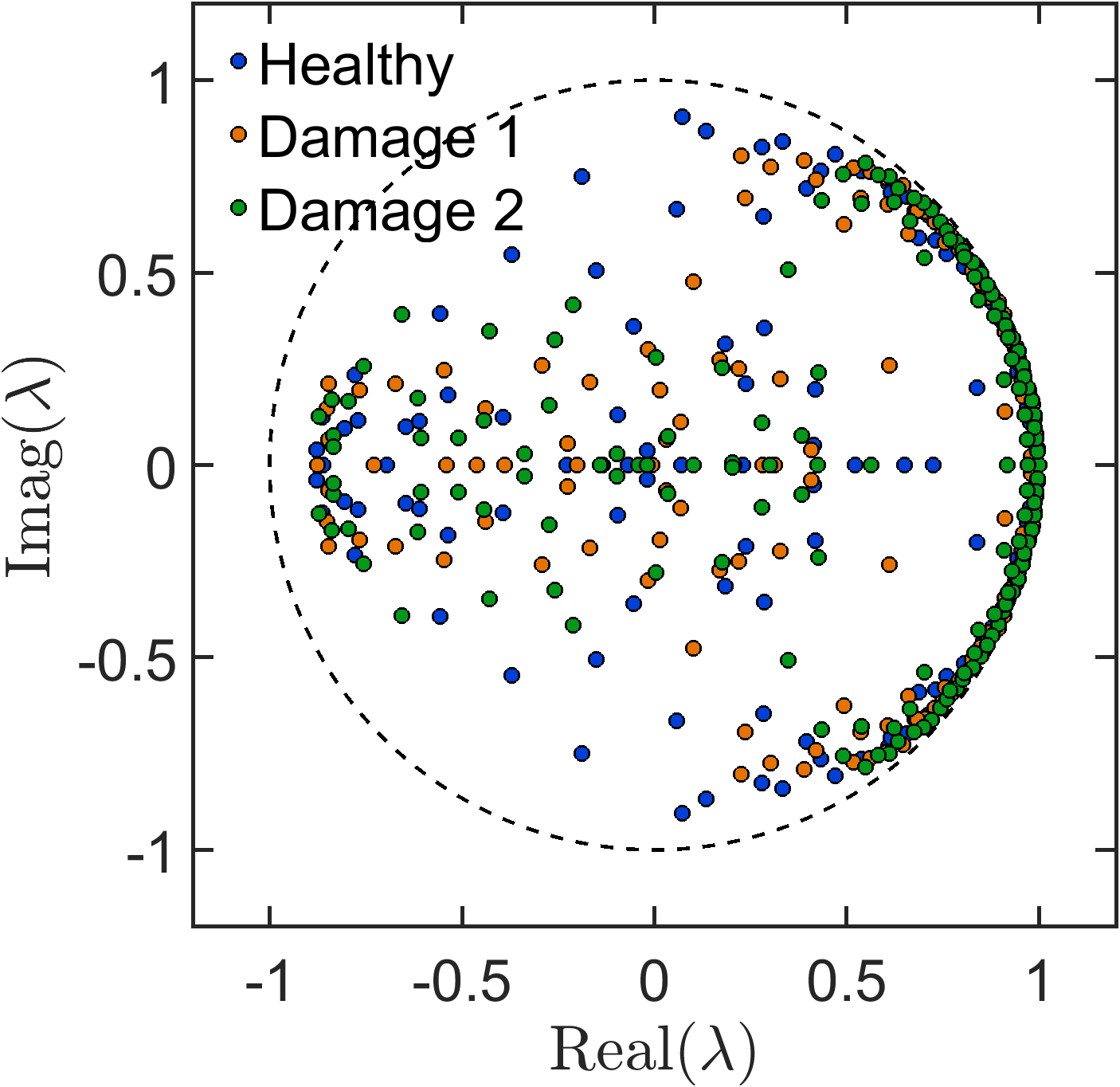}}
    \put(230,20){\includegraphics[width=0.48\columnwidth]{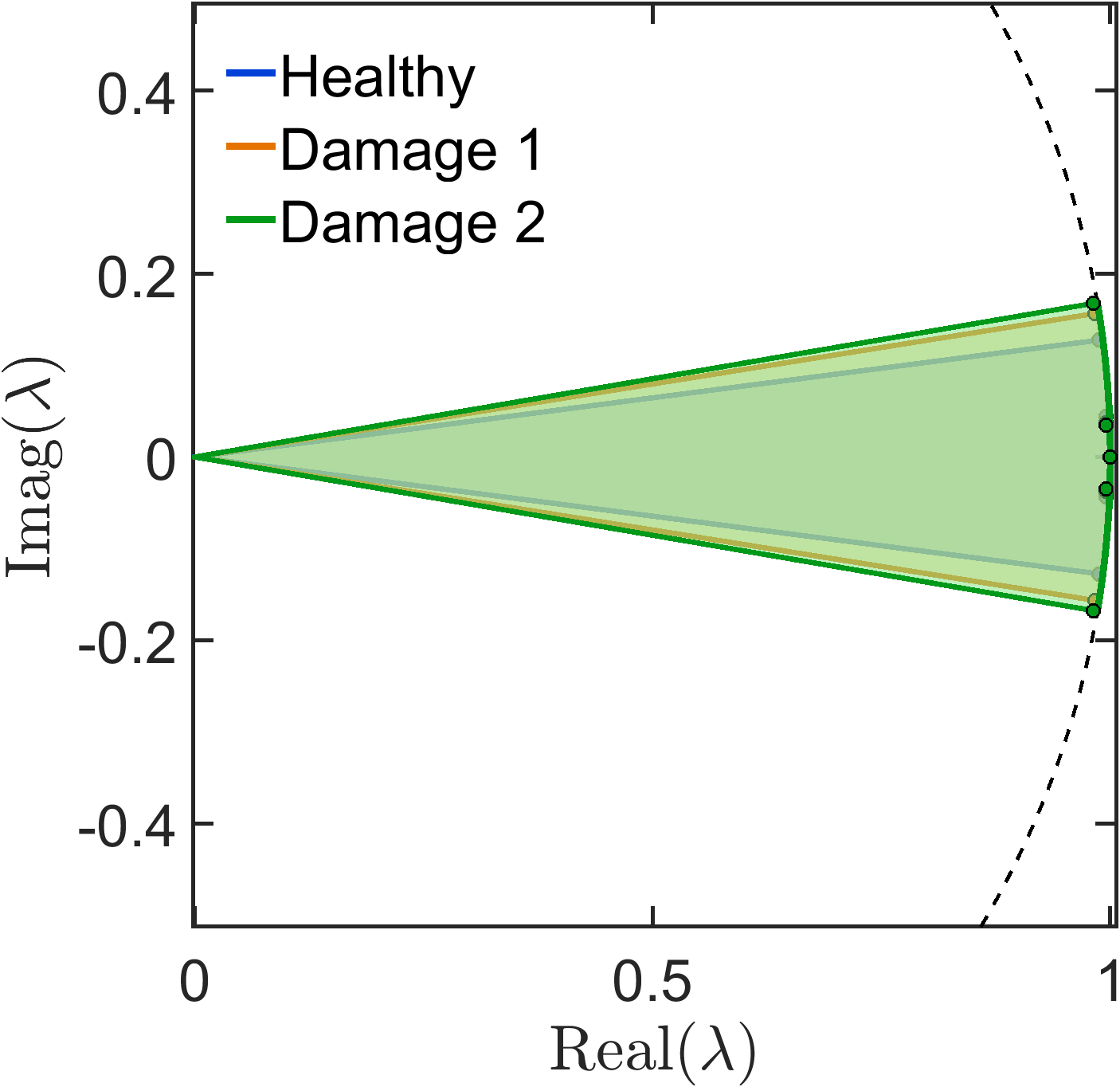}}

    \put(20,250){\large \textbf{(a)}}
    \put(280,250){\large \textbf{(b)}}
\end{picture}

\vspace{-0.3cm}
\caption{(a) Eigenvalues obtained from the DMD algorithm for different structural states; and (b) enclosed area formed by the first five dominant eigenvalues for different structural states.}
\label{fig:eigenvalue_area_combined}
\end{figure}

\begin{figure}[H]

\begin{picture}(400,180)
    \put(90,0){\includegraphics[width=0.6\columnwidth]{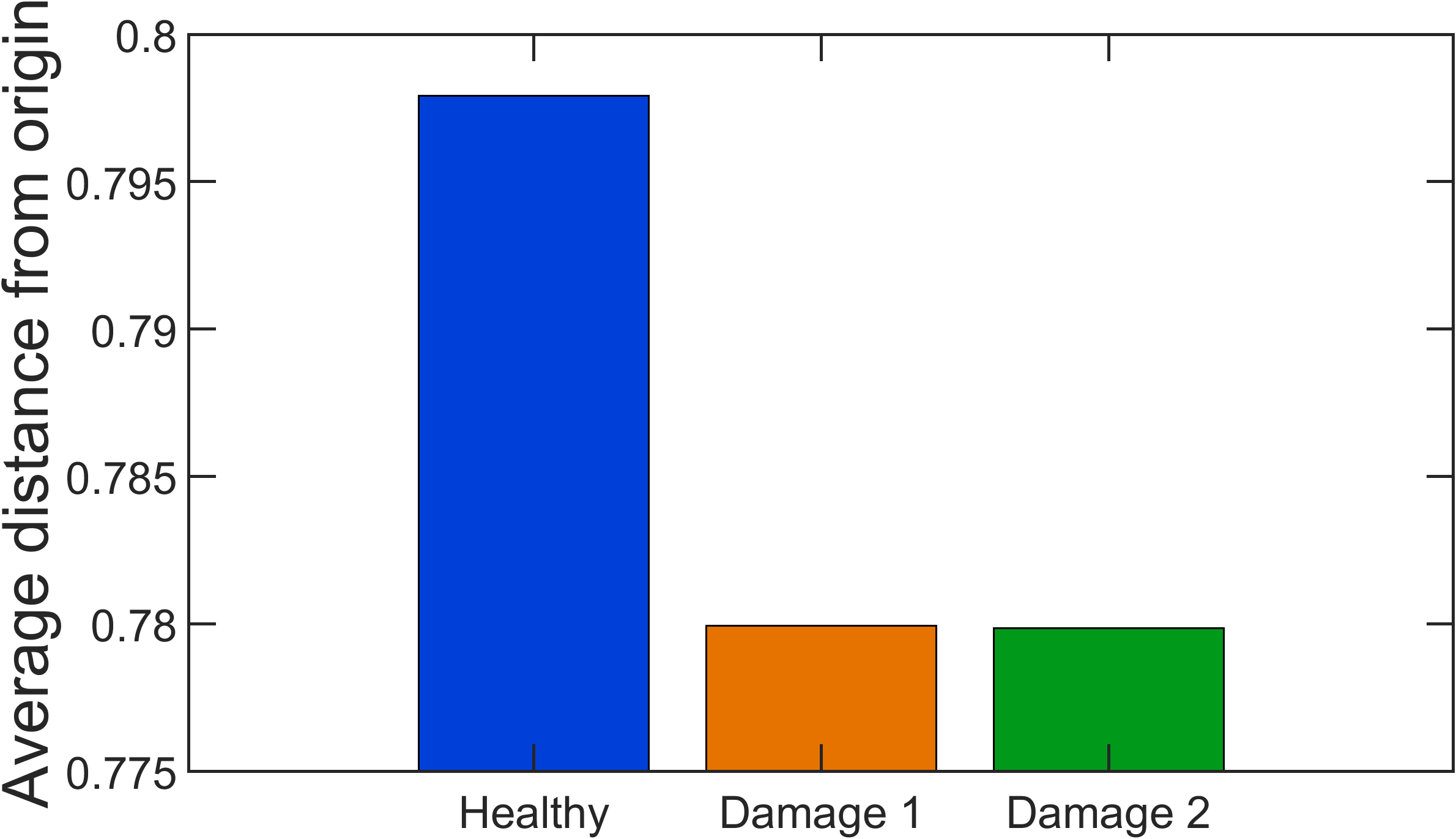}}
\end{picture}

\vspace{0.2cm}\caption{Average distance from origin of the eigenvalues.}
\label{fig:avg_distance_origin}
\end{figure}

Figure~\ref{fig:eigenvalue_area_combined} presents the distribution of DMD eigenvalues for the healthy and damaged configurations, along with the regions enclosed by the leading dominant eigenvalues. As observed, the eigenvalues corresponding to the damaged cases exhibit a slight inward shift toward the origin relative to the healthy configuration. In a discrete-time setting, the magnitude of each eigenvalue governs the rate of temporal decay; therefore, this inward shift indicates increased damping or faster attenuation of the identified dynamics. This behavior is further supported by the average distance-from-origin metric shown in Figure~\ref{fig:avg_distance_origin}, where the healthy case exhibits the highest mean eigenvalue magnitude, while both damaged cases show consistently lower values. Since eigenvalues closer to the unit circle correspond to more weakly decaying (i.e., more persistent) dynamics, the observed inward shift for the damaged configurations suggests that structural degradation is associated with enhanced energy dissipation in the measured response. A similar trend is observed in the regions enclosed by the first five dominant eigenvalues, as shown in Figure~\ref{fig:eigenvalue_area_combined}(b). The damaged configurations exhibit larger enclosed areas compared to the healthy case, indicating a greater dispersion of dominant eigenvalues in the complex plane.

\begin{figure}[htb]

\begin{picture}(400,380)
    % top row
    \put(-10,180){\includegraphics[width=0.5\columnwidth]{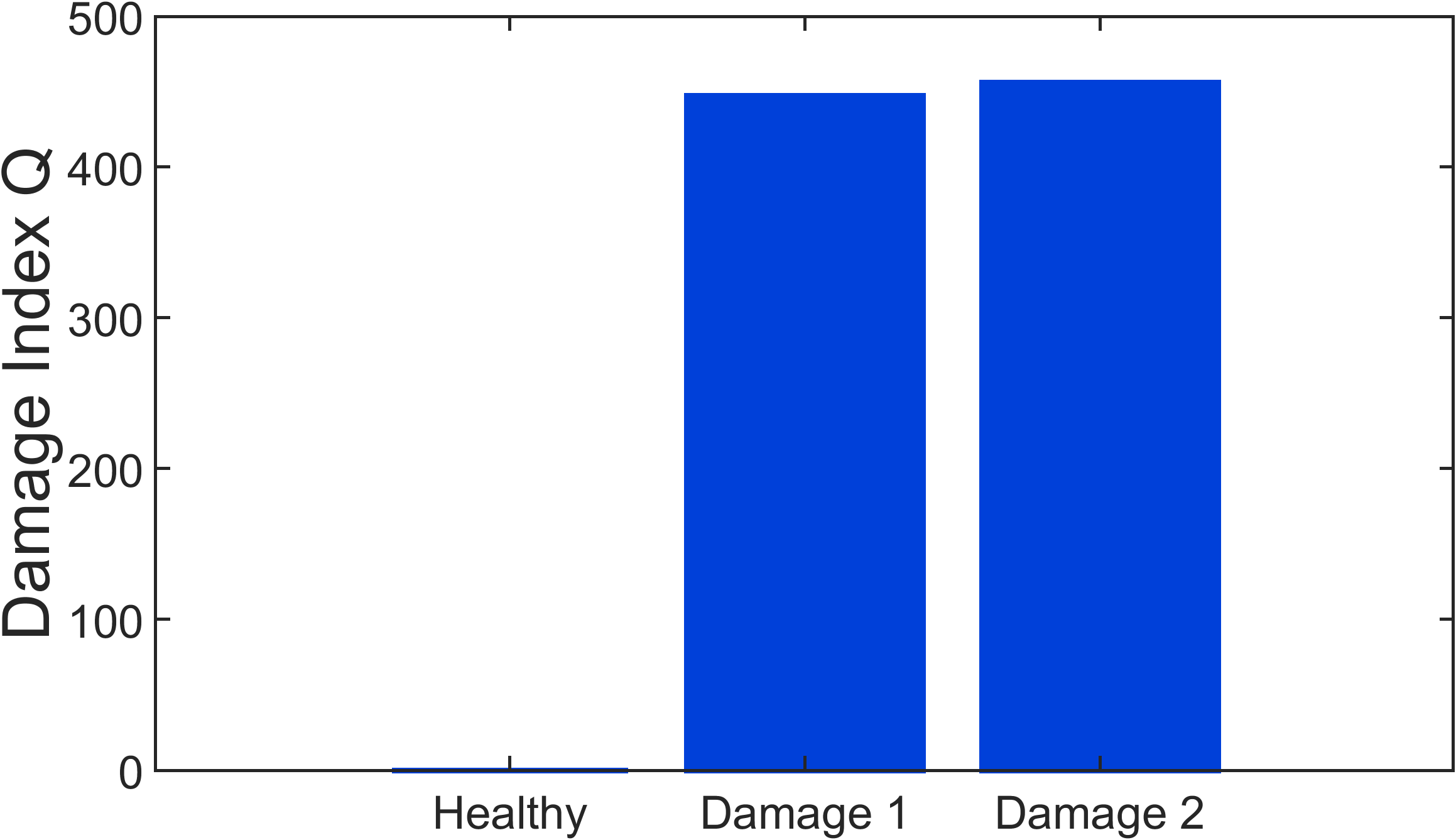}}
    \put(230,180){\includegraphics[width=0.5\columnwidth]{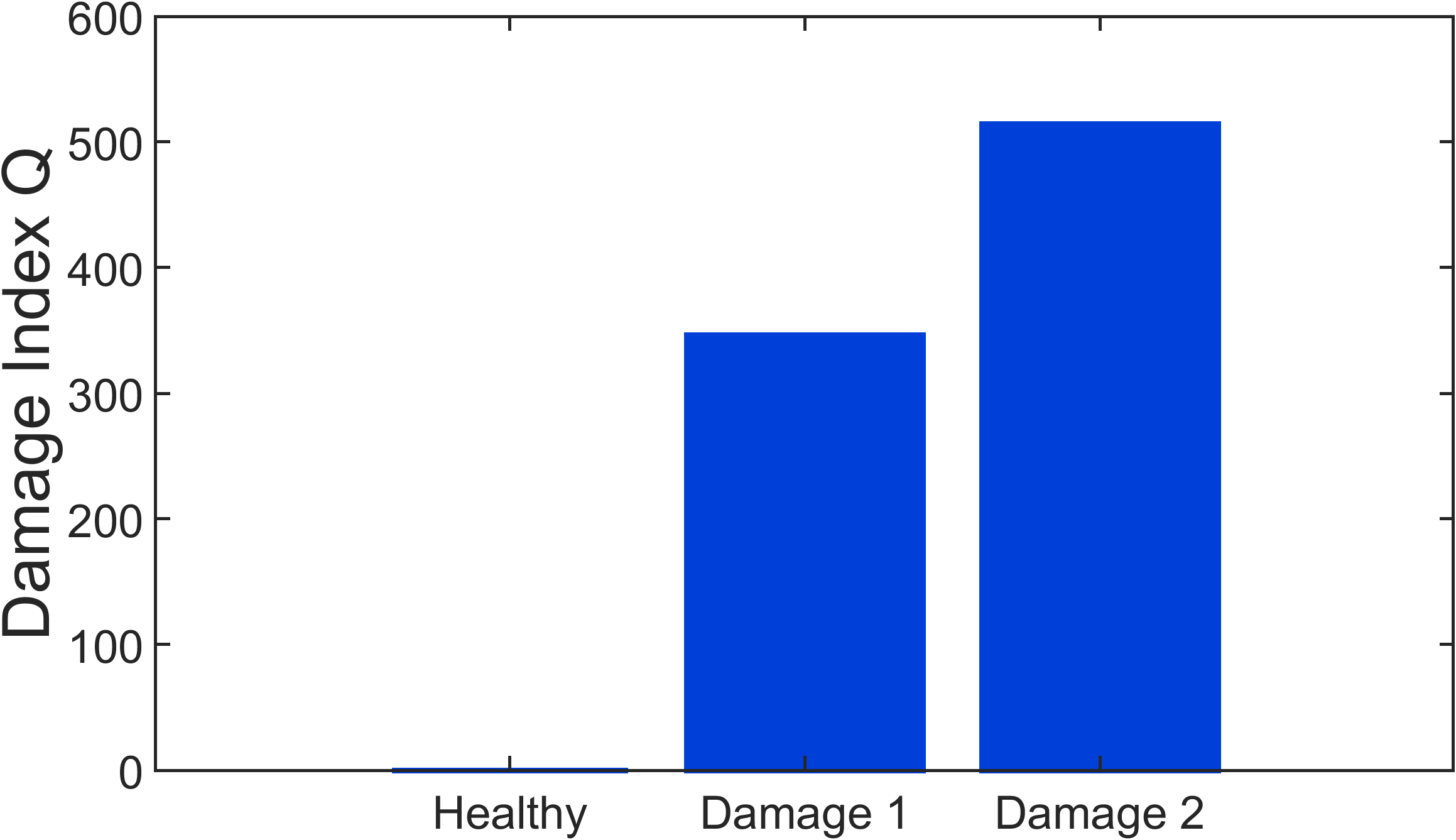}}

    % bottom row
    \put(-10,0){\includegraphics[width=0.5\columnwidth]{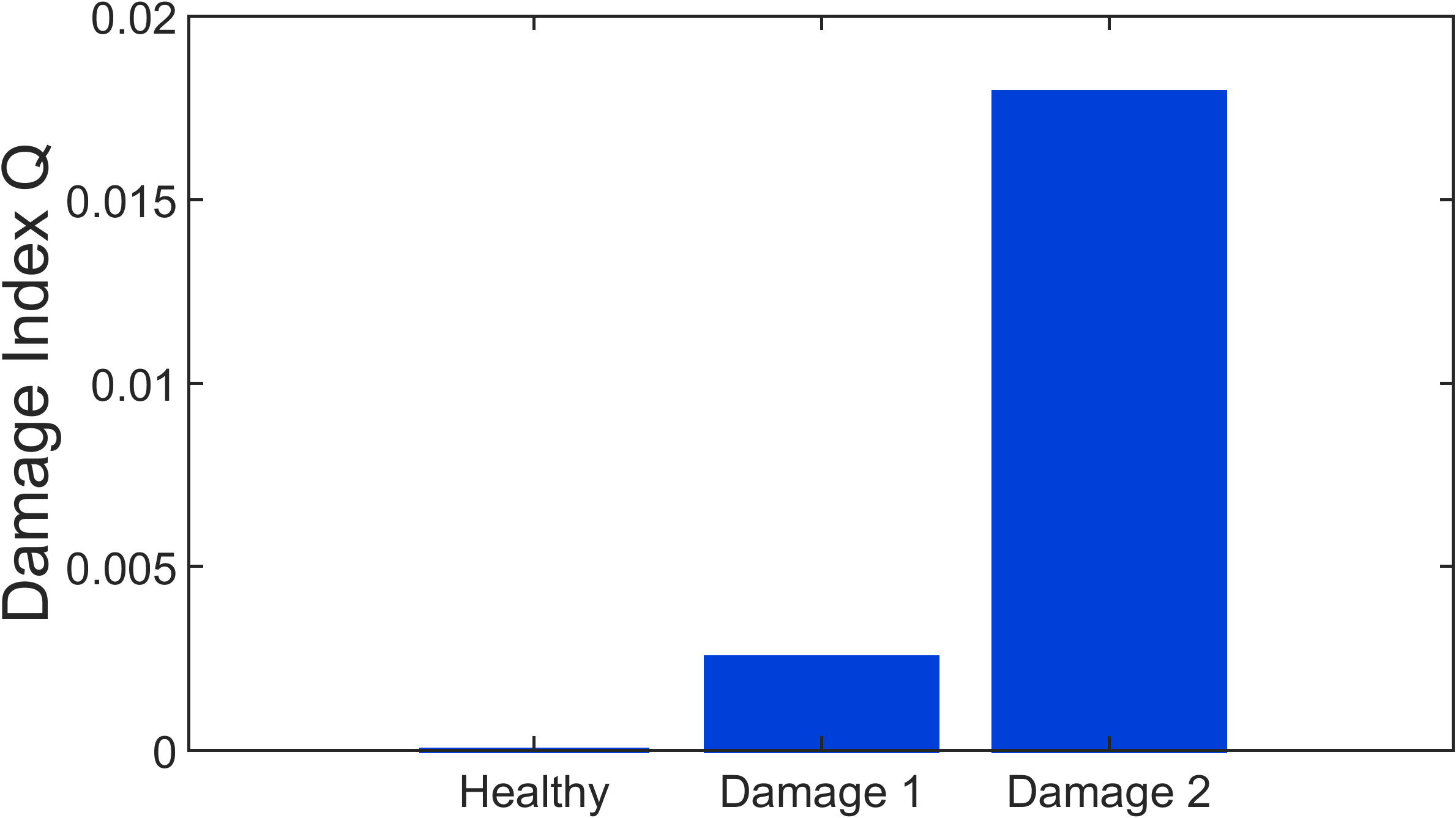}}
    \put(230,0){\includegraphics[width=0.5\columnwidth]{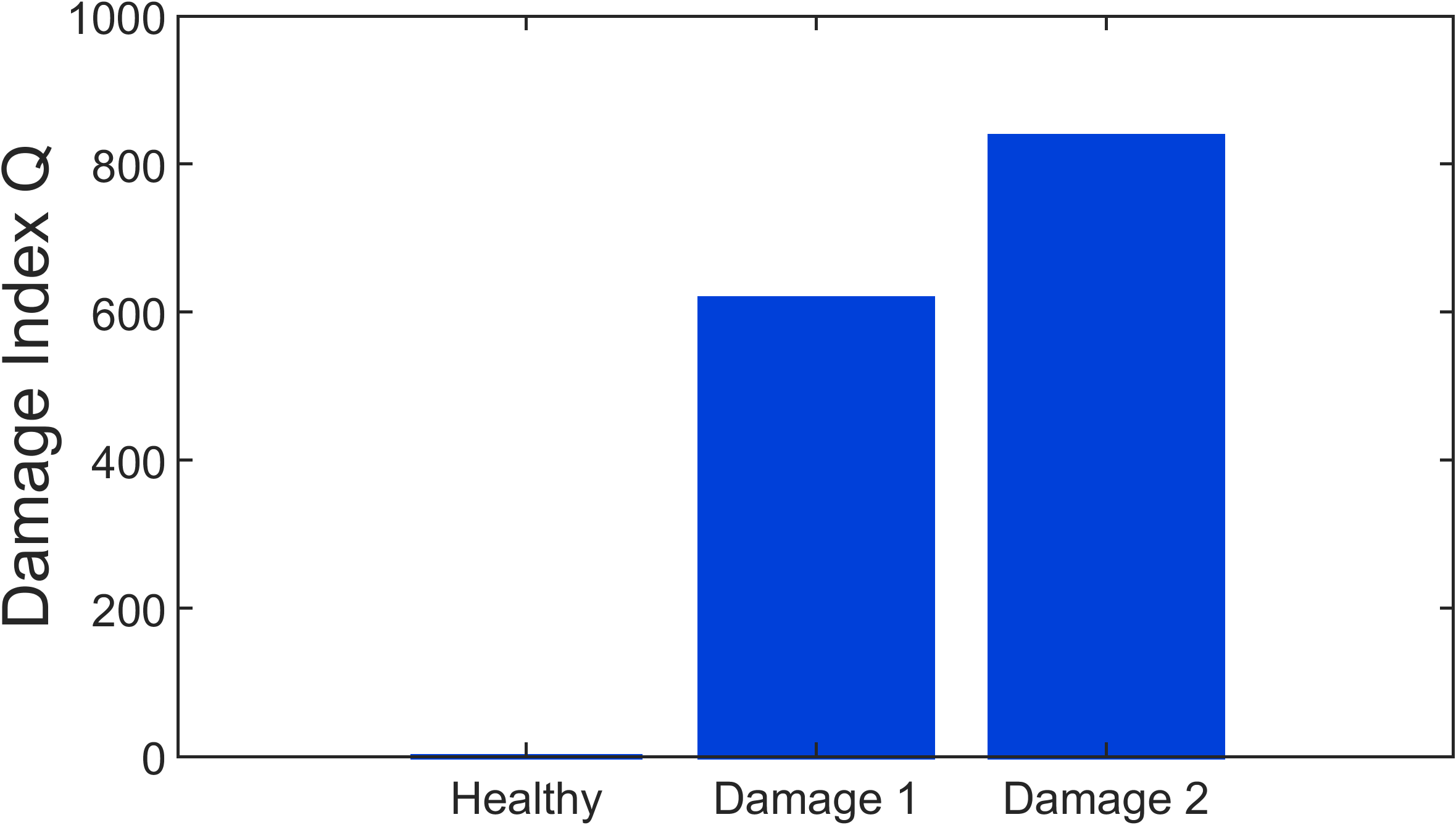}}

    % labels
    \put(20,330){\large \textbf{(a)}}
    \put(280,330){\large \textbf{(b)}}
    \put(20,150){\large \textbf{(c)}}
    \put(280,150){\large \textbf{(d)}}
\end{picture}

\vspace{0.3cm}\caption{Damage index $Q$ for different structural states based on different modal characteristics: (a) Mode Shape (MS); (b) Mode Shape Curvature (MSC); (c) Mode Shape Curvature Square (MSCS); and (d) Mode Shape Slope (MSS).}
\label{fig:damage_indices}
\end{figure}
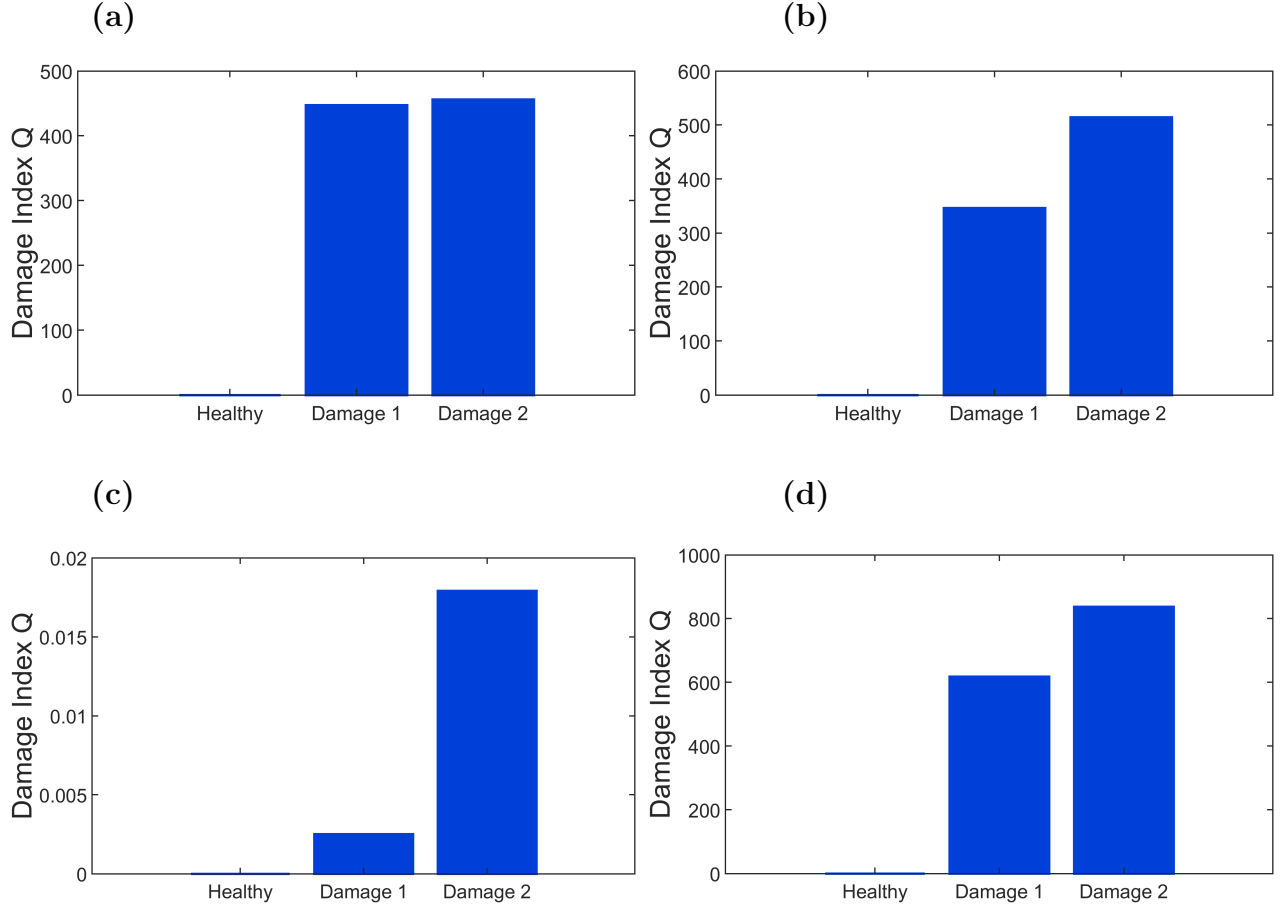

 A damage sensitive feature defined as $Q$ based on the modal features---including mode shape (MS), mode shape slope (MSS), mode shape curvature (MSC), and mode shape curvature square (MSCS)---were constructed. The corresponding damage indices are presented in Figure~\ref{fig:damage_indices}. As expected, the healthy configuration yields a zero value of $Q$, serving as the reference state, while the damaged configurations exhibit significantly higher values across all feature definitions. Furthermore, a consistent increase in $Q$ from Damage 1 to Damage 2 is observed for all indices, indicating that the proposed formulation effectively captures both the presence and progression of structural damage. The clear separation between healthy and damaged states across all subplots demonstrates the robustness of the $Q$-based metric for damage detection.

However, their sensitivity to damage severity is not identical. The MS based index shows only a slight increase from Damage 1 to Damage 2, suggesting limited sensitivity to  damage severity once the presence of damage has already been established. In contrast, the MSC and MSS based indices exhibit a more noticeable increase, while the MSCS-based index shows the largest proportional change between the two damaged states. Although the absolute magnitude of the MSCS based index remains smaller than the other features, its  relative increase indicates a higher sensitivity to damage severity. This enhanced sensitivity of the MSCS based feature can be attributed to its dependence on the squared mode shape curvature. Structural damage causes a reduction in stiffness, which increases the curvature in the damaged region. When this curvature term is squared, the variation is amplified further, so even moderate curvature differences between Damage 1 and Damage 2 can produce a substantial increase in the corresponding damage index. 

Overall, these results demonstrate that the index $Q$, when constructed from DMD-derived modal features, provides a physically interpretable measure for distinguishing healthy and damaged structural conditions.

\section{Conclusion}

This study demonstrates that Dynamic Mode Decomposition (DMD) provides a robust and effective data-driven framework for structural damage identification using both simulation and experimental datasets. The results consistently show that DMD accurately reconstructs and predicts system dynamics, with low errors indicating that the identified linear representation reliably captures the dominant behavior even beyond the training horizon. 

Analysis of the extracted DMD modes reveals that structural damage induces measurable changes in the spatial characteristics of the system, while the eigenvalue distributions provide complementary insight into the temporal dynamics. In particular, the observed inward shift and increased dispersion of eigenvalues in the damaged cases indicate enhanced attenuation associated with structural degradation. 

Furthermore, the proposed damage index $Q$, constructed from DMD-derived modal features, enables clear separation between healthy and damaged states and demonstrates consistent sensitivity to damage severity across all cases. 

Overall, these findings highlight that the integration of DMD with modal feature extraction offers a unified, physically interpretable, and computationally efficient approach for vibration-based structural health monitoring, enabling both accurate dynamic representation and reliable damage detection without the need for extensive preprocessing.

\section{Limitations and Future Work}

Although the proposed framework demonstrates strong capability in distinguishing between healthy and damaged structural states, it is currently limited to damage detection and quantification of the severity of damage. However, it does not perform damage localization. Future work will focus on extending the framework to enable damage localization. Additionally, the methodology will be further validated and adapted for more complex and higher-dimensional structural systems, including those with nonlinear behavior.

%%%%% Acknowledgments %%%%%%%%%%%%%%%%%%%%%%%%%%%

\section*{Acknowledgments}

We acknowledge the start-up funding support provided by the department of Mechanical Engineering at South Dakota State University.

\bibliographystyle{unsrt} 
\bibliography{ref}
 \end{document}